\documentstyle[prb,aps,epsf]{revtex}

\begin{document}

\draft

\title{How to escape Aharonov-Bohm cages~?}

\author{Julien Vidal$^{1}$,  Patrick Butaud $^{2}$, Benoit Dou{\c 
c}ot$^{3,4}$, and R\'emy
Mosseri$^{1}$}

\address{$^1$ Groupe de Physique des Solides, CNRS UMR 7588,
Universit\'{e}s Paris 6 et 7,\\
2, place Jussieu, 75251 Paris Cedex 05 France}

\address{$^2$ Centre de Recherches sur les Tr\`es Basses
Temp\'eratures, Laboratoire CNRS associ\'e \`a l'Universit\'e Joseph Fourier\\
25, avenue des Martyrs, 38042 Grenoble France}

\address{$^3$ Laboratoire de Physique Th\'{e}orique et
Hautes \'Energies, CNRS UMR 7589, Universit\'{e}s Paris 6 et 7,\\
4, place Jussieu, 75252 Paris Cedex 05 France}

\address{$^4$ Laboratoire de Physique de la Mati\`ere Condens\'ee, 
CNRS UMR 8551,
\'Ecole Normale Sup\'erieure,\\
24, rue Lhomond, 75231 Paris Cedex 05 France}

\maketitle

\begin{abstract}

We study the effect of disorder and interactions on a recently 
proposed magnetic field
induced localization mechanism. We show that both partially destroy 
the extreme confinement of
the excitations occuring in the pure case and give rise to unusual 
behavior. We also point out the role
of the edge states that allows for a propagation of the electrons in 
these systems.
 
\end{abstract}

\pacs{PACS numbers: 72.15.Rn, 73.20.Dx.}

%
%
%

\section{Introduction}
\label{Introduction}

%
%
The dynamics of a quantum particle in a periodic potential in the 
presence of a uniform static
magnetic field has revealed many beautiful effects and has been the 
subject of ongoing researches
for several decades. The competition between the two characteristic 
length scales involved, namely
the lattice period and the magnetic length  produces a very complex 
pattern of energy levels. In
two dimensions, this leads to fractal structures in the spectrum anticipated by
Azbel\cite{Azbel}, and further studied by 
Hofstadter\cite{Hofstadter}, Claro and
Wannier\cite{Claro_Wannier}, Rammal\cite{Rammal_hexa}, and many others.
Some striking experi\-mental manifestations of these effects have 
been found in macroscopic
properties of two-dimensional superconducting wire networks, such as 
the variation of the
critical temperature with respect to the external magnetic field, the 
magnetization or the critical
current. The connection between the linearized Ginzburg-Landau 
equations used to describe these
wire networks close to the superconducting transition, and the 
tight-binding spectrum of
electrons had been established slightly before these experiments by 
De Gennes\cite{DeGennes_fils_supra} and
Alexander\cite{Alexander_fils_supra}.

More recently, a surprising effect has been presented in 
ref.~\cite{Vidal_Cages} in a
two-dimensional lattice with hexagonal symmetry, the so-called ${\cal 
T}_{3}$ lattice displayed in
Fig.~\ref{Star3lattice}. For some values of the magnetic field 
corresponding to half a flux quantum per elementary
tile, the energy spectrum  of a tight-binding model with nearest 
neighbor hopping collapses into three highly
degenerate levels. Furthermore, we have shown that it is possible to 
build energy eigenstates where the probability to
find an electron is non vanishing only in a finite size cluster which 
we have called an Aharonov-Bohm cage. This
corresponds to a localization mechanism due to  quantum interferences 
of Aharonov-Bohm type between paths enclosing a
half-integer number of flux quanta. For instance, if an electron is 
initially located on a given site of the lattice,
it will never propagate beyond the boundary of the cage associated to 
this site, provided the magnetic field is chosen
properly, as discussed above.

Since this theoretical observation, several experiments have searched 
for manifestations of these cages on real systems. A
serie of investigations on superconducting wire networks has been 
completed by Abilio {\it et al.}\cite{Abilio_T3}
and by Serret {\it et al.}\cite{Serret_T3} providing a good agreement 
between predictions and  measurements of the
critical temperature and the critical current.  In addition, the 
unusual nature of the mixed
state for half a flux quantum per elementary tile has been clearly 
indicated by a strong reduction of the critical
current, and by magnetic decoration experiments which shows a very 
highly disordered vortex
pattern\cite{Serret_T3,Pannetier_vortex_T3}.

A slightly more direct probe of this localization effect has been 
provided by transport measurements in mesoscopic lattices
carved in a two-dimensional electron gas at the interface of 
GaAs/GaAlAs heterostructures\cite{Naud_T3}. In theses
systems, the number of tranverse conduction channels in the wires can 
be very small (a few units), and the mean free path
much larger than the distance between two nearest neighbor nodes. At 
intermediate magnetic fields, Naud {\it et al.}
have clearly observed a periodic modulation of the magnetoconductance 
with a period of one flux quantum per plaquette
($h/e$). In the same experimental conditions, a square lattice does 
not show these oscillations. It seems more than
plausible that these
$h/e$-periodic oscillations seen on ${\cal T}_{3}$ lattice could be 
attributed to the Aharonov-Bohm cages. Indeed, this
localization mechanism establishes a strong difference in the 
tranport properties between integer and half integer fluxes,
which is expected to survive  in the presence of a weak enough 
disorder as discussed in ref.~\cite{Vidal_Transmission}.

These experiments clearly motivate us to address the question of the 
robustness of these cages in the
presence of ``real life" perturbations such as disorder, finite size 
effects and electron-electron interactions. The main
purpose of this paper is to attempt to bridge the gap between the 
very idealized tight-binding model initially studied in
ref.~\cite{Vidal_Cages}, and the experimental systems mentioned 
above. As a general trend, we will show that if the
cages are in themselves very fragile, their existence in the ideal 
system induces a few remarkable features for the
perturbed ones. The underlying reason for this is that these 
perturbations act on a very degenerate system, and as
often in physics, degenerate perturbations are expected to induce 
some very interesting phenomena. One of the most
famous example is, of course, the fractional quantum Hall effect 
where the degeneracy of the kinetic energy at
fractional filling factors opens the way to the formation of strongly 
correlated many-body exotic states such as the
Laughlin liquid\cite{Laughlin_liquid}.

This paper is organized as follows. Section \ref{Model} presents a 
detailed discussion of the butterfly-like
energy spectrum for a tight-binding model on the ${\cal T}_{3}$ 
lattice. Section \ref{Cages} focusses on half integer
fluxes for which the cages appears. Several viewpoints are given to 
provide a simple physical intuition of what happens
in the system. This enables us to show other examples of lattices for 
which Aharonov-Bohm cages exist. Section
\ref{Disorder} discusses disorder effects still within the 
tight-binding framework. So, it is complementary to the
previous study presented in ref.~\cite{Vidal_Transmission} where the 
continuous wire model was analyzed, in
connection to the experiments on two-dimensional gases. Here, we 
investigate some properties of one particle eigenstates
such as the inverse participation ratio as a function of the disorder 
strength and the magnetic field.
Section \ref{Edges} is devoted to edge states. These states are found 
to be confined in a finite width strip along the
boundaries of the sample. We outline the subtleties involved in the 
translation from the tight-binding model properties to
those of the continuous one-dimensional wire networks. Finally, 
section \ref{Interactions} adresses interactions
effects in the context of the Hubbard model. Most of the results deal 
with the two electron case. In agreement with a
previous study for the chain of loops\cite{Vidal_Chapelet}, we find 
that interactions are able to create some extended
and dispersive two particle eigenstates. This can be viewed as a kind 
of interaction induced delocalization phenomenon,
similar to the one discussed by Shepelyansky for disordered 
systems\cite{Shepelyansky}. The
presence of the cages in the non interacting case is reflected by the 
fact that these extended states involve a close
binding of the two particles in real space, even if the local 
interaction is repulsive. We also give some ground state
configurations for a finite density of particles and we show that the 
energy per particle has a singular behaviour when
the number of electrons per site becomes larger than 1/3.
In the last section, we conclude and discuss the experimental 
relevance of our work. Technical details
for the two interacting particles problem can be found in the 
appendix \ref{Diago}.

%
%
%

\section{Spectral properties}
\label{Model}

%
%
%

\begin{figure}
\centerline{\epsfxsize=95mm
\epsffile{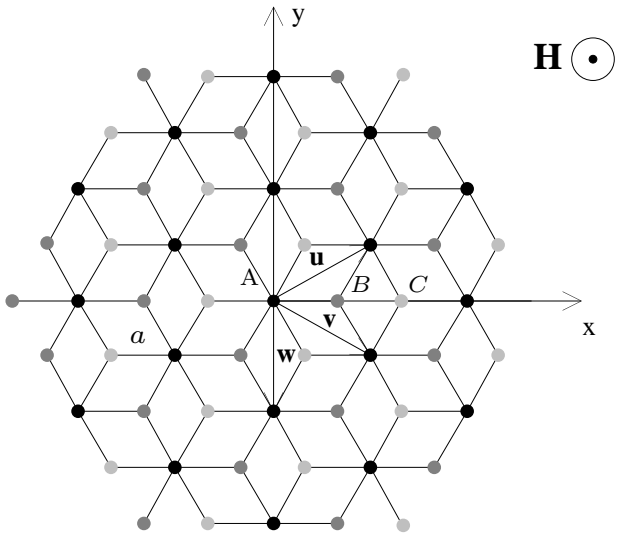}}
\caption{A piece of the ${\cal T}_{3}$ lattice embedded in the 
perpendicular magnetic field
${\bf H}$~; $a$ denotes the lattice spacing.}
\label{Star3lattice}
\end{figure}

We consider the ${\cal T}_{3}$ lattice displayed in 
Fig.~\ref{Star3lattice}  which is a bipartite periodic structure with
3 sites per unit cell: one 6-fold coordinated site $A$ and two 3-fold 
coordinated site $B$ and $C$.
%
%
%
%
This tiling can also be seen as the dual of the most famous Kagom\'e 
lattice\cite{Baxter}.
We consider a tight-binding hamiltonian defined by~:
%
%
\begin{equation}
H_0=\sum_{<i,j>}t_{ij}|i\rangle \langle j|
\label{hamil}
\end{equation}
%
%
where $|i\rangle$ is a localized orbital on site $i$.
When ${\bf H}=0$, the hopping term $t_{ij}=1$ if $i$ and $j$ are nearest
neighbors and $0$ otherwise.
In the presence of a magnetic field\cite{Peierls},
$t_{ij}$ is multiplied by a phase factor $e^{i \gamma _{ij}}$ involving the
vector potential ${\bf {A}}$~:
%
%
\begin{equation}
\gamma _{ij}={\frac{2\pi }{\phi _{0}}}\int_{i}^{j}{\bf A}.\mbox{d}{\bf l}
\mbox{,}
\end{equation}
%
%
where $\phi _{0}=hc/e$ is the flux quantum. In the following, we 
focus on the case of a
uniform magnetic field ${\bf H}=H{\bf z}$ which can be obtained, for 
example, with the
Landau gauge ${\bf A}=H(0,x,0)$, and we denote by $\phi 
=Ba^{2}\sqrt{3}/2$ the magnetic
flux through an elementary rhombus.

\begin{figure}[h]
\centerline{\epsfxsize=100mm
\epsffile{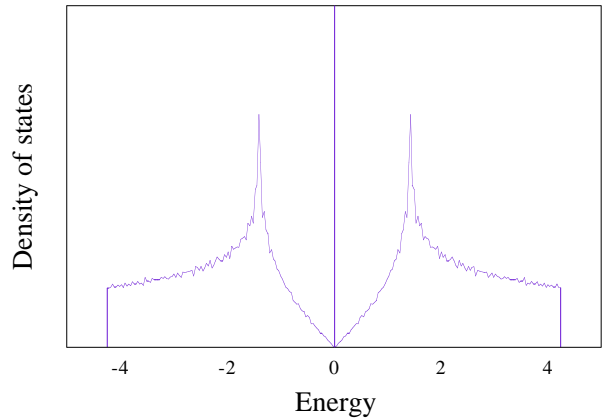}}
\vspace{-4mm}
\caption{Density of states of the ${\cal T}_{3}$ lattice for ${\bf H}=0$.}
\label{DOST3}
\end{figure}

Let us first briefly analyze the zero field case ${\bf H}=0$. In this case, the
one-particle spectrum is given by the following dispersion relation~:
%
%
\begin{equation}
E_{\pm}\left({\bf k}\right)=\pm
\sqrt{6+4\left[\cos({\bf k}.{\bf u})+\cos({\bf k}.{\bf v})
+\cos({\bf k}.{\bf w})\right]}
\mbox{,}
\end{equation}
%
%
where ${\bf u}=a (3/2,\sqrt{3}/2)$ and ${\bf v}=a (3/2,-\sqrt{3}/2)$ 
are the vectors of
the primitive cell (see Fig.~\ref{Star3lattice}), ${\bf w}={\bf 
v-u}$, and where
${\bf k}$ is a wave vector lying in the first Brillouin zone.
The associated eigenstates are standard Bloch waves.
In addition, there is a non dispersive band at $\varepsilon_0=0$
resulting from the bipartite character of the structure, and whose 
degeneracy is given by
the difference between the number of 3-fold and 6-fold coordinated 
sites. Since its origin is purely
topological, this energy will always be an eigenvalue for any ${\bf 
H}$, with the same degeneracy.
The density of states, originally computed by 
Sutherland\cite{Sutherland_T3}, is shown in
Fig.~\ref{DOST3}.
%
%
%
%

\begin{figure}[h]
\centerline{\epsfxsize=100mm
\epsffile{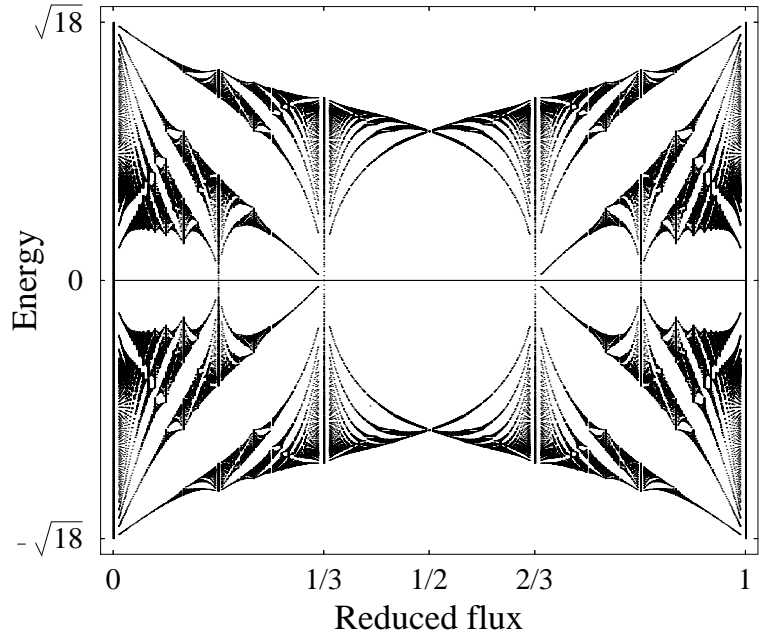}}
\caption{Spectrum of the ${\cal T}_{3}$ lattice as a function of the 
reduced flux.}
\label{butterfly}
\end{figure}

For ${\bf H} \neq 0$, the spectrum is of course much more complex 
since there remains only
one  translational invariance  along the $y$ direction due to the 
gauge choice, so that the
eigenfunctions can be written as~:
%
%
\begin{equation}
\varphi_j (x,y)=\psi_j(x)e^{ik_{y}y},\hspace{1ex}k_{y}\in \left[
0,2\pi /{a\sqrt{3}}\right]
\mbox{.}
\end{equation}
%
where the index $j= A,B,C$. Thus, the secular equations read~:
%
%
\begin{eqnarray}
\label{eq1}
\varepsilon \psi_A (x)&=&\psi_B(x+a)+\psi_B\left(x-{a\over2}\right)
2\cos\left[{\gamma\over a} \left(x-{a\over4}\right)+\kappa\right]+ \nonumber \\
& &\psi_C(x-a)+\psi_C\left(x+{a\over2}\right)2\cos\left[{\gamma\over a}
\left(x+{a\over4}\right)+\kappa\right] \\
\nonumber\\
\label{eq2}
\varepsilon \psi_B(x)&=&\psi_A(x-a)+\psi_A\left(x+{a\over2}\right)
2\cos\left[{\gamma\over a}\left(x+{a\over4}\right)+\kappa\right] \\
\nonumber\\
\label{eq3}
\varepsilon \psi_C(x)&=&\psi_A(x+a)+\psi_A\left(x-{a\over2}\right)
2\cos\left[{\gamma\over a}\left(x-{a\over4}\right)+\kappa\right]
\mbox{ , }
\end{eqnarray}
%
%
where $\kappa=k_y a \sqrt{3}/2$, and $\gamma=2 \pi \phi/ \phi _{0}=2 
\pi f$. Note that with
the origin $O$ chosen in Fig.~\ref{Star3lattice}, $x$ only takes 
integer or half-integer values.
If one looks for solutions $\varepsilon\neq 0$ of the secular system, one can
substitute (\ref{eq2}) and (\ref{eq3}) in (\ref{eq1}) to obtain an effective
one-dimensional equation  that only involves six-fold coordinated sites~:
%
%
\begin{eqnarray}
(\varepsilon ^{2}-6)\psi _{m}&=&2\cos\left({\frac{\gamma}{2}}\right)
\left\{ 2\cos\left[{\frac{3\gamma}{2}}\left(m+{\frac{1}{2}}\right)
+\kappa\right]\psi_{m+1}\right.    +
\left.2\cos\left[{\frac{3\gamma}{2}}\left(m-{\frac{1}{2}}\right)
+\kappa \right]\psi_{m-1}\right. +2\cos\left(3\gamma 
m+2\kappa\right)\psi_{m}\bigg\}
\mbox{ , }
\label{spectre1d}
\end{eqnarray}
%
%
where $\psi_{m}=\psi_A({x=3ma/2})$, $(m\in {\bf Z})$. As it can be 
readily seen in
Fig.~\ref{Star3lattice}, the six-fold coordinated sites form a 
triangular lattice so
that Eq.~(\ref{spectre1d}) has to be compared to those derived by Claro and
Wannier for the triangular lattice\cite{Claro_Wannier}~:
%
%
\begin{eqnarray}
\varepsilon _{T}\psi _{m} &=&2\cos \left[ \gamma _{T}\left( m+{\frac{1}{2}}
\right) +\kappa \right] \psi _{m+1} +2\cos \left[ \gamma _{T}\left( 
m-{\frac{1}{2}}\right)
+\kappa \right] \psi _{m-1} +2\cos (2\gamma _{T}m+2\kappa )\psi _{m}
\end{eqnarray}
%
%
with similar notations. So, the ${\cal T}_{3}$ lattice spectrum can 
be simply obtained from the
triangular lattice spectrum  by setting $\gamma _{T}=3\gamma /2$, 
since one has~:
%
%
\begin{equation}
\varepsilon ^{2}-6=2\cos \left( {\frac{\gamma }{2}}\right) \varepsilon _{T}
\mbox {.}
\label{homo}
\end{equation}
%
%
Since the tiling is bipartite, the spectrum is symmetric ($\varepsilon
\leftrightarrow -\varepsilon $). Moreover, Eq. ($\ref{spectre1d}$) 
displays a translation
symmetry $f\rightarrow f+n\: (\forall n\in {\bf Z})$ and a reflection 
invariance about
half-integer values of $f$.
We thus limit our analysis to $0 \leq f \leq 1/2$. For rational 
values of $f=p/q$ ($p,q$
mutually prime), the system ($\ref{spectre1d}$) becomes closed after 
a translation by $q$
periods and the spectrum is made up of $2q$ bands. Note that, when 
$q=3q^{\prime }$
$(q^{\prime }\in {\bf N})$, this period is actually reduced by a 
factor $3$ and the spectrum
has only $2q^{\prime }$ bands.
In particular, for $f=1/3$, one can exactly compute the dispersion relation~:
%
%
\begin{equation}
\varepsilon^2(\kappa,k_x)-6=-2\cos(2\kappa) +4\cos(\kappa) \cos(k_x a)
\mbox{ , }
\end{equation}
%
%
where $(k_x,\kappa) \in [0,2\pi/a\sqrt{3}] \times [0,\pi]$.
For this flux, the spectrum spreads from $-3$ to $+3$ and is gapless.
Figure \ref{butterfly} shows the ${\cal T}_{3}$ lattice spectrum 
support versus the reduced flux $f$.
The most spectacular and unusual feature is that, for $f=1/2$, the 
spectrum collapses into
three eigenvalues $\varepsilon_0 =0$ and $\varepsilon_\pm =\pm 
\sqrt{6}$ as it can be
readily seen from  Eq.~(\ref{homo}).
%
%
%
%

This is all the more curious that usually, for an infinite periodic 
structure and for rational
values of $f$, the spectrum is absolutely continuous (band-like). 
Here, it behaves as a
{\it super-atom} with three infinitely degenerate levels with equal 
spectral weights $1/3$. The
existence of these non dispersive bands suggests the possibility to 
build Wannier-type  localized
eigenstates. This remarkable fact is, for any electron density, 
susceptible to induce original
physical properties  reminiscent of those of a localized system.

%
%
%

\section{The Aharonov-Bohm cages}
\label{Cages}

%
%
%

In this section, we shall analyze these properties from the quantum 
dynamics point of view by
characterizing the spreading of a wave packet in the ${\cal T}_{3}$ 
lattice, at $f=1/2$. To achieve this,  it
is worth focusing on the spectral charateristics at a local level. 
The magnetic field being
uniform, the whole spectrum is indeed recovered from the local 
density of states (LDOS) on the two
different types of sites.  At this point, it is more convenient to 
shift from the above Landau
gauge to a cylindrical (symmetric) gauge defined by ${\bf 
A}=H(-y/2,x/2,0)$ that respects the
discrete rotational symmetry existing locally.  We proceed to an 
analytic Lanczos tridiagonalization
using the recursion  method\cite{Haydock_rec1}. The
principle of this algorithm is to generate an semi-infinite chain 
according to the
recursive relation~:
%
%
\begin{equation}
b_{n+1}|\varphi_{n+1} \rangle =(H_0-a_n)|\varphi_n \rangle - b_n 
|\varphi_{n-1} \rangle
\label{recurs}
\end{equation}
%
%
which allows to evaluate the LDOS on the initial orbital $|\varphi_0 
\rangle$. The diagonal and
off-diagonal terms  ($a_{n}$,$b_{n}$) are obtained by requiring the 
normalization of the orbitals
$|\varphi_n\rangle$. Note that in our case, all the diagonal elements 
$a_{n}$ vanish because~:
({\it i}) the structure is bipartite, ({\it ii}) $H_0$ is purely off-diagonal.

\begin{figure}[h]
\centerline{\epsfxsize=80mm
\epsfbox{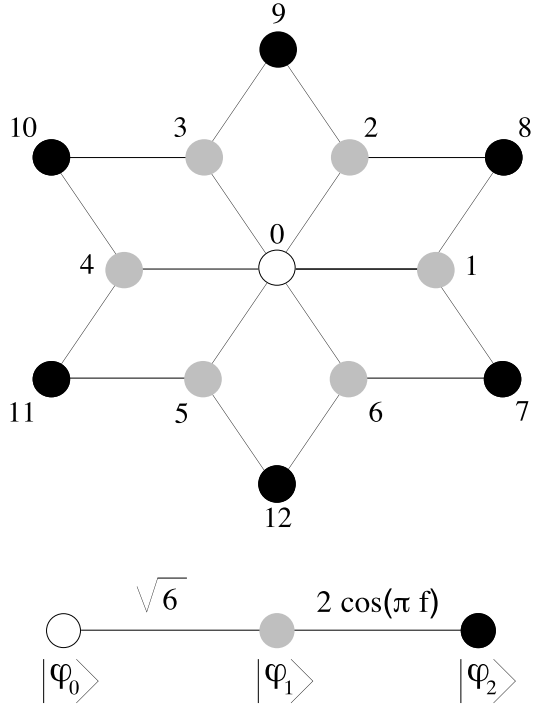}}
\vspace{-20mm}
\caption{
Aharonov-Bohm cage associated to the six-fold coordinated sites and 
its recursion chain.
For $f=1/2$, the black sites are never visited by a wave packet 
initially localized on the
central site (white).}
\label{Recursion6}
\end{figure}

Let us first compute the recursion chain associated to a six-fold 
coordinated site and choose, as
initial orbital, $|\varphi_0 \rangle=|0\rangle$ (see 
Fig~.\ref{Recursion6}). One has~:
%
%
\begin{equation}
H_0|\varphi_0\rangle=\sum_{i=1}^6 |i\rangle
\hspace{3ex} \Rightarrow \hspace{3ex}
  b_1=\sqrt{6} \, , \:
|\varphi_1\rangle={1 \over \sqrt{6}}\sum_{i=1}^6 |i\rangle
\mbox{ . }
\end{equation}
%
%
The first recursion orbital $|\varphi_1\rangle$ is a linear symmetric 
combination of the first shell
sites (three-fold coordinated grey sites). At the next step, one obtains~:
%
%
\begin{equation}
H_0|\varphi_1\rangle=b_1 |\varphi_0\rangle + 2\cos(\pi f)
{1 \over \sqrt{6}}\sum_{i=7}^{12} |i\rangle
\hspace{3ex} \Rightarrow \hspace{3ex}
b_2=2\cos(\pi f)  \, , \:|\varphi_2\rangle={1 \over 
\sqrt{6}}\sum_{i=7}^{12} |i\rangle
\mbox{ . }
\end{equation}
%
%
%
%
%
%
For $f=1/2$, $b_2=0$ so that the LDOS is reduced to two values 
$\varepsilon_\pm =\pm \sqrt{6}$ with
an equal spectral weight $1/2$. This effect can be simply understood 
in terms of Aharonov-Bohm effect.
Indeed, the amplitude of probability ${\cal A}_{0 \rightarrow 2 
\rightarrow 9}$  to go for example, ``in two steps", from
  0 to 9 {\it via} 2, is exactly the opposite of ${\cal A}_{0 
\rightarrow 3 \rightarrow 9}$, so that the
resulting amplitude is zero. Then, any wave packet initially 
localized on a six-fold coordinated site
($|\psi(t=0)\rangle=|0\rangle$) is completely trapped inside what we 
have called an Aharonov-Bohm cage\cite{Vidal_Cages}
whose precise definition is given below.  One can easily determine 
its time evolution that is periodic and given by~:
%
%
\begin{equation}
|\psi(t)\rangle=\cos\left(\sqrt{6}t\right) |\varphi_0\rangle - i 
\sin\left(\sqrt{6}t\right) |\varphi_1\rangle
\mbox{ . }
\end{equation}
%
%
In particular, the autocorrelation function $|\langle 0|\psi (t)\rangle|^2=
\cos^2\left(\sqrt{6}t\right)$ for $f=1/2$, whereas for $f=0$~:
%
%
\begin{equation}
|\langle 0|\psi (t)\rangle|^2=\left({3\sqrt{3}\over 8\pi^2 a^2} \int\int_{1BZ}
\cos\left( E\left({\bf k}\right)t\right) {\mbox d}{\bf k}
\right)^2
\end{equation}
%
%
($1BZ$ stands for first Brillouin zone) and behaves at large $t$ as~:
%
%
\begin{equation}
|\langle 0|\psi (t)\rangle|^2\simeq
{3\over 8 \pi^2 t^2}
\left[\sin\left(3\sqrt{2} t\right)+\sqrt{3}\cos\left(\sqrt{2} 
t\right) \right] ^2
\mbox{ . }
\end{equation}
%
%

We emphasize that for a generic rational $f\neq 1/2$, the hamiltonian 
is periodic and one expects
$|\langle 0|\psi(t)\rangle|^2\sim 1/t^2$. {\it A contrario},  when 
$f$ is irrationnal, the hamiltonian is
quasiperiodic and one expects, as for the square lattice \cite{KPG}, 
$|\langle 0|\psi(t)\rangle|^2\sim
1/t^{2\alpha}$ with $\alpha<1$ since the spectrum should be singular 
continuous.

\begin{figure}[h]
\centerline{\epsfxsize=80mm
\epsfbox{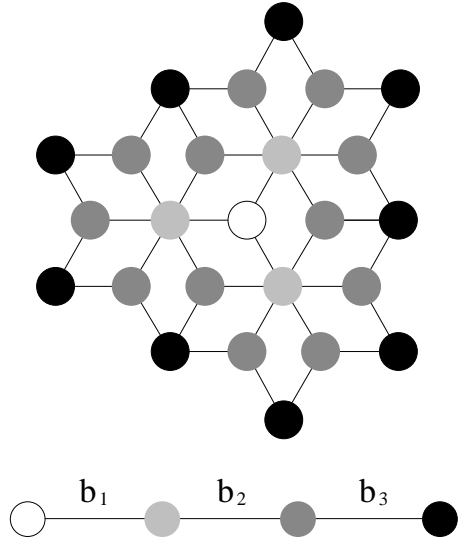}}
\caption{\label{recur3}
Aharonov-Bohm cage associated to the three-fold coordinated sites and 
its recursion chain.}
\end{figure}

The same type of recursion calculation can be made for the three-fold 
coordinated sites. In this case, one
easily obtains~:
%
%
\begin{eqnarray}
b_1&=&\sqrt{3} \, , \\
b_2&=&\sqrt{4 \cos^2(\pi f) + 3} \, , \\
b_3&=&\sqrt{8\cos^2(\pi f) + 4\left[\cos(\pi f) + \cos(3 \pi f)\right]^2 \over
4 \cos^2(\pi f) + 3}
\mbox{ , }
\end{eqnarray}
%
%
so that $b_3=0$ for $f=1/2$, and the LDOS on this type of site reduces to the
$\varepsilon_0=0$ (spectral weight 1/2), and 
$\varepsilon_\pm=\pm\sqrt{6}$ (spectral weight 1/4).
The quantum evolution of a wave packet initially localized on such a 
site is therefore, as previously,
confined in a cage displayed in Fig.~\ref{recur3} that is larger than 
the one obtained for the 6-fold
coordinated sites.
%
%
%
%
Finally, since the ${\cal T}_{3}$ lattice is only made up of 6-fold 
and 3-fold coordinated sites,
any wave packet (with a finite initial extension) will be confined 
inside a bound Aharonov-Bohm
cage  at $f=1/2$.  We emphasize that this localization phenomenon 
must be understood from the
quantum dynamics point of view. Indeed, contrary to the Anderson 
localization, the eigenstates of
the system are not exponentially localized, and not even localized at 
all since the high
degeneracy allows one to build any type of eigenstates (possibly extended).\\

To proceed further, it is necessary to give a precise definition of 
these Aharonov-Bohm cages.
For a given wave packet $|\psi\rangle$ submitted to a magnetic field 
characterized by a reduced flux
$f$, we define the Aharonov-Bohm cage ${\cal C}^f_{\psi}$ as the  set 
of sites visited by this
initial wave packet during its evolution. In general,  ${\cal 
C}^f_{\psi}$ is infinite, but as shown
previously, it can, for specific values of the magnetic field 
($f=1/2$  for the ${\cal T}_{3}$ lattice), be
bound. To analyze the cage structures for any independent electron 
model, it is sufficient to
characterize the cages of all inequivalent sites.  Indeed, the superposition
principle implies that if $|\psi(t=0)\rangle=\sum_{i=1}^n \alpha_i 
\,|i\rangle$,
one has the following property~:
%
%
\begin{equation}
{\cal C}_{\psi}^f \:{ \subset} \: \bigcup_{i=1}^n \:{\cal C}_{i}^f
\mbox{ . }
\end{equation}
%
%

\begin{figure}[h]
\centerline{\epsfxsize=130mm
\epsfbox{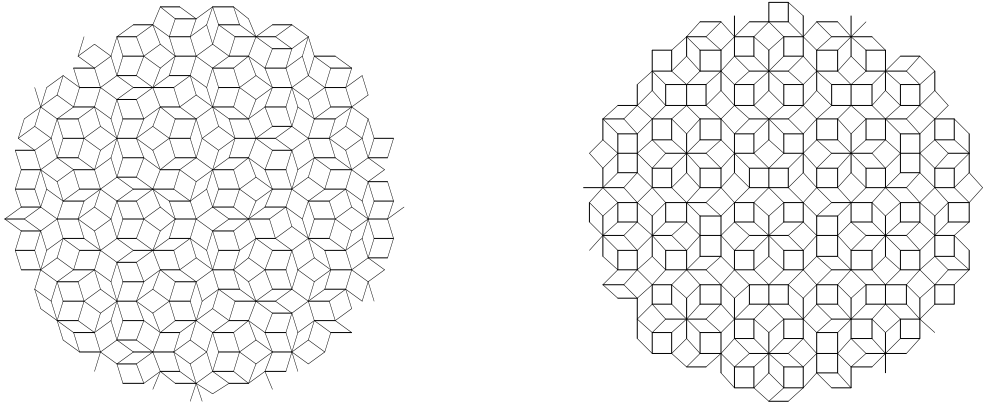}}
\vspace{5mm}
\caption{Two examples of structure presenting bound cages for special 
initial conditions of the
wave packet localized at the center of some 5-fold symmetric stars 
for the Penrose tiling (left)
and 8-fold symmetric stars for  the octogonal tiling (right).}
\label{PenOcto}
\end{figure}

With this definition, several cases can be encountered~:
\begin{itemize}

\item all  cages are unbound for any $f$ (ex~: square lattice, 
triangular lattice, honeycomb)~;

\item some cages are bound for particular values of $f$ (ex~: the 
Penrose tiling and the
octagonal  tiling displayed in Fig.~\ref{PenOcto})~;

\item all the cages are bound for the same values of $f$ (ex~: ${\cal 
T}_{3}$ lattice).

\end{itemize}

%
%
%
%

\begin{figure}[h]
\centerline{\epsfxsize=120mm
\epsfbox{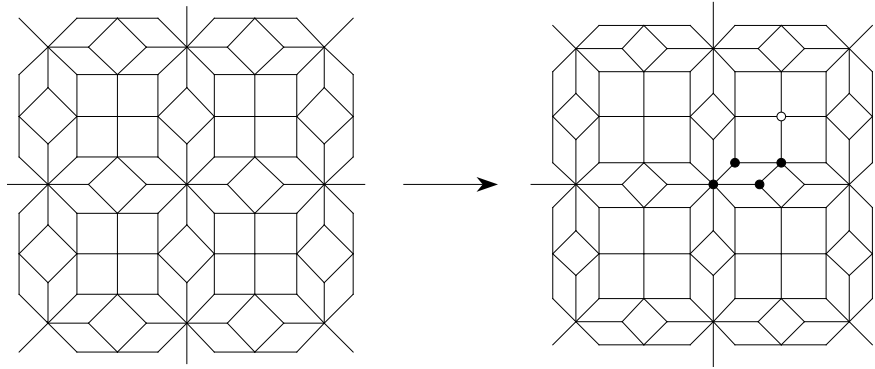}}
\caption{
A piece of the periodic ${\cal T}_{4}$ tiling before (left) and after 
deformation (right). The five
different types of sites are represented ($\bullet, \circ$). }
\label{Star4lattice}
\end{figure}

\begin{figure}[h]
\centerline{\epsfxsize=50mm
\epsfbox{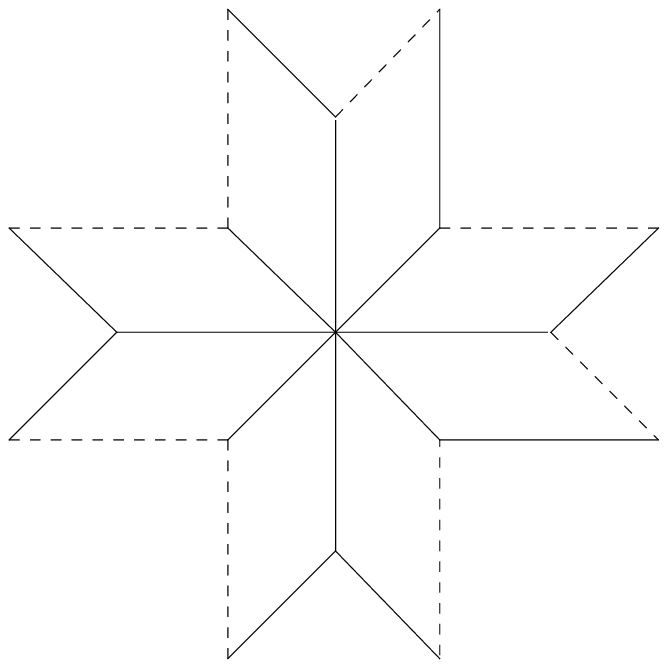}}
\caption{
A possible gauge choice for the ${\cal T}_{4}$ lattice at $f=1/2$. 
All the Peierls terms are equal to
$+1$ except those represented by dotted lines that are equal to $-1$.
For clarity we have omit the hopping term $t_{L,S}$.}
\label{Star4gauge}
\end{figure}

This latter case, that we will call fully confined structure, can be 
met for other
tilings. As an example, we have displayed in Fig.~\ref{Star4lattice} 
, the so-called
${\cal T}_{4}$ lattice which can be obtained by deforming an 
approximant with $14$ sites
per unit cell of the octagonal tiling.
%
%
%
%
This structure has exactly the same physical properties as the ${\cal 
T}_{3}$ lattice at $f=1/2$
($f$ refers here to the reduced flux measured in the smallest tile). 
Thereafter,  we shall consider
a {\it light} version of this tiling obtained by eliminating the 
sites $\circ$  and their related bonds (see
Fig.~\ref{Star4lattice}). The resulting structure is simply a set of 
connected 8-fold
symmetric stars with two different lengths whose ratio is equal to 
$\sqrt{2}$. We thus allow for
two different hopping terms
$t_L$ and $t_S$ for long and short edges respectively. For $f=1/2$, 
it is always possible to choose a gauge such
that all the $t_{ij}$'s are real. A possible choice is represented in 
Fig.~\ref{Star4gauge}.
%
%
%
%
Note that this gauge has the same periodicity as the structure. This is quite
surprising since it is not the generic case (see next section for the 
${\cal T}_{3}$ lattice).
It is then easy to diagonalize $H_0$ and to obtain the thirteen 
following eigenvalues~: 0
(three-fold degenerate), $\pm \sqrt{2 t_L^2+2 t_S^2}$  (two-fold 
degenerate), $\pm \sqrt{t_S}, \pm \sqrt{4
t_L^2+4 t_S^2},$ and $\pm \sqrt{4 t_L^2+2 t_S^2}$. For any hopping 
terms $t_{L,S}$ these eigenvalues do not
depend on any wave vector and thus form the exact analogous of the 
non dispersive bands
obtained in the ${\cal T}_{3}$ lattice.

\begin{figure}[h]
\centerline{\epsfxsize=100mm
\epsfbox{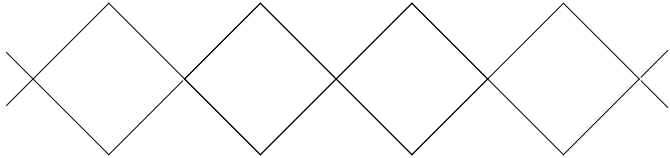}}
\caption{
The simplest example of a periodic fully confined structure for $f=1/2$.}
\label{chap}
\end{figure}

We would like to point out that fully confined systems can also be 
obtained in one dimension. We have shown in
Fig.~\ref{chap}  a quasi-one-dimensional structure that clearly 
displays bound Aharonov-Bohm cages at $f=1/2$. Its
eigenspectrum is, as for the ${\cal T}_{3}$ lattice, made up of three 
non dispersive bands with energies $0,\pm 2$.
%
%
%
%

After this detailed description of this magnetic field induced 
localization, it is natural to wonder if
this phenomenon is robust to various type of ``perturbations". For 
example, a non uniform magnetic field
has been shown to drastically change the energy spectrum and the 
localization properties of the system providing a
complete destruction of the cages\cite{Oh_T3}. In the next section, 
we study the effect of different kind of disorder
and we show that, the destructive interference that leads to the 
Aharonov-Bohm cages are partially destroyed.

%
%
%

\section{The disorder}
\label{Disorder}

%
%
%

\begin{figure}[h]
\centerline{\epsfxsize=60mm
\epsfbox{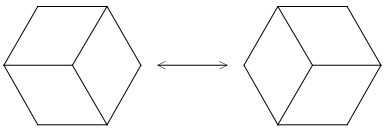}}
\vspace{5mm}
\caption{Elementary flip for two-dimensional tilings.}
\label{flip}
\end{figure}

A first natural way to introduce disorder in a system is to directly 
modify its structure by incorporating
defects. In the case of the ${\cal T}_{3}$ lattice, such defects can 
be simply generated by a local {\it elementary
flip} of three tiles as shown in Fig.~\ref{flip}.
%
%
%
%
For $f=1/2$, it is clear that a finite number of flips does not 
modify the bound nature of the
cages  since, sooner or later, a given wave packet will be embedded 
in the ``pure"  ${\cal T}_{3}$ lattice
geometry which is responsible of the confinement. Nevertheless, one 
can wonder if this feature
still holds for a finite density of such defects.  Indeed, if we 
consider a large
number of flips in order to get closer and closer to a random tiling
configuration\cite{Elser_random_tiling}, we expect the destructive 
Aharonov-Bohm
interferences to completely disappear. It would be, in this case, 
interesting to determine the
critical density below which all the cages remains bound. This could 
be achieved by considering an extended
initial state and by computing its time evolution for different 
realizations of the disorder and
for different flip densities. Note that this value should be easily 
understood in terms of cages percolation
threshold.\\

Another possible way to introduce disorder is to randomly modulate 
either off-diagonal or diagonal
terms in the hamiltonian. In this study, we mainly focus on the 
latter case and consider the following
hamiltonian~:
%
%
\begin{equation}
H_{dis.}=\sum_{\langle i,j \rangle}t_{ij}\,|i\rangle \langle j| +
\sum_i V_i \, |i\rangle \langle i|
\label{ham.Anderson}
\mbox{ , }
\end{equation}
%
%
where the $t_{ij}$ are defined as in (\ref{hamil}), and where the 
on-site energies $V_i$ are
mutually independent gaussian random variable with variance $\sigma^2$.

As the strength of disorder increases, the strongly localized cage 
eigenstates begin to spread over larger
distances. In some sense, a weak disorder increases the one-particle 
localization typical length~! However,
since the system is two-dimensional, we expect the overall 
localization length to be finite for any value of
$\sigma \neq 0$ and for any energy. For a given reduced flux $f$, as 
$\sigma$ is increased from zero, the
one-particle eigenstates first get mixed within the energy bands of 
the pure model. This regime corresponds
to the standard Anderson localization. This lasts until $\sigma$ 
becomes of the order of some energy gaps
allowing then for interband mixing.

For $f=1/2$, the remarkable structure of the spectrum at zero disorder is
responsible for the absence of the first disorder induced 
localization regime. Therefore, we expect, in this
case, that most observables will depend only weakly on $\sigma$, as 
long as $\sigma$ remains smaller than
the gaps. To analyze this problem, it is interesting to construct an 
effective model in the lowest energy
subspace of
$H_0$, expressed in the localized cage basis (see 
Fig.~\ref{cagebasis} in the next section). Denoting the
normalized cage eigenstates associated to
$\varepsilon_-=-\sqrt{6}$ centered around the six-fold coordinated 
sites ${\bf r}$ by $|{\bf r}\rangle$,
this effective hamiltonian has the form~:
%
%
\begin{equation}
H_{eff.}=\sum_{\langle {\bf r} ,{\bf r}' \rangle} \tilde t_{{\bf r},{\bf r}'}\,
|{\bf r}\rangle \langle {\bf r}'| +
\sum_{\bf r} \tilde V_{\bf r} \, |{\bf r}\rangle \langle {\bf r}|
\mbox{ , }
\end{equation}
%
%
where $\langle {\bf r} ,{\bf r}' \rangle$ stands for the sum over 
nearest neighbors in the triangular
lattice generated by the the six-fold coordinated sites. The 
effective parameters
$\tilde t_{{\bf r},{\bf r}'}$ and $\tilde V_{\bf r}$ depend on the 
$V_j$'s. For simplicity, we use the
indexation of Fig.~\ref{Recursion6} so that~:
%
%
\begin{eqnarray}
\tilde V_{\bf 0}&=&-\sqrt{6}+{V_0 \over 2}+{1\over 12} \sum_{i=1}^{6} V_i \\
\tilde t_{{\bf 0,8}}&=&{1\over 12} (V_1-V_2) e^{i \alpha_{0,8}}
\mbox{ , }
\end{eqnarray}
%
%
where $\alpha_{0,8}$ is a gauge dependent phase factor. The general 
patterns for $\tilde V_{\bf r}$ and
$\tilde t_{{\bf r},{\bf r}'}$ can be easily inferred from these 
examples. We observe that $H_{eff.}$
involves both diagonal and off-diagonal disorder with comparable 
magnitude since~:
$\langle \tilde V_{\bf r}^2\rangle-\langle \tilde V_{\bf 
r}\rangle^2=7 \sigma^2/24$ and
$\langle |\tilde t_{{\bf r},{\bf r}'}|^2\rangle-\langle |\tilde 
t_{{\bf r},{\bf r}'}|\rangle^2=
\sigma^2/72$. It is interesting to note that non vanishing tunneling 
amplitude from one cage to a neighboring
one are induced by an asymmetry between the two diagonal energies of 
the intermediate 3-fold coordinated
sites connecting the cage centers. According to this effective model, 
the statistical properties of
the eigenstates and  the energy levels are independent of $\sigma$, 
provided the eigenenergies  are
rescaled $\varepsilon \rightarrow (\varepsilon+\sqrt{6})/ \sigma$.

\begin{figure}[h]
\centerline{\epsfxsize=85mm
\epsfbox{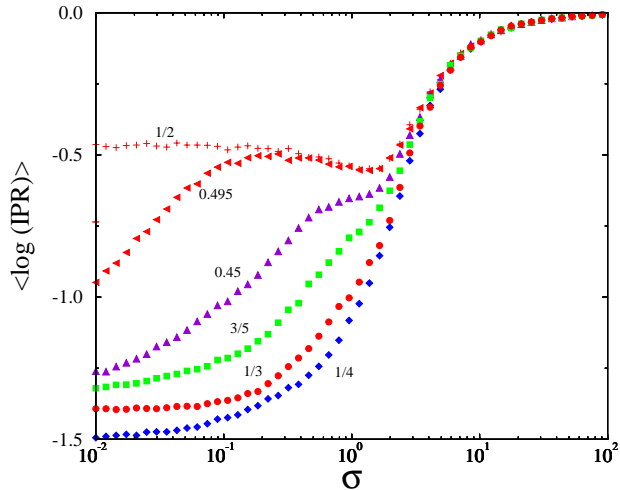}}
\caption{Variation of the averaged $\log$ (IPR) as a function of the 
disorder strength $\sigma$, for several
values  of $f$ ($N_s=605$ sites). The average is done over energies 
and over disorder realizations.}
\label{IPR}
\end{figure}

We have checked this simple picture by numerical diagonalization of 
finite clusters of size as large as 605
sites with open boundary conditions. As discussed in the next 
section, this introduces edge states. Here,
the shape of the clusters has been properly chosen to minimize the 
number of these eigenstates whose energy, in
the pure system at $f=1/2$, differs from $0, \pm \sqrt{6}$. To 
characterize the degree of localization of a
given eigenstate $|\psi \rangle$, it is convenient to compute the 
inverse participation ratio (IPR) defined
by~:
%
%
\begin{equation}
\mbox{IPR}(\psi)={\sum_i |\psi_i|^4 \over \left( \sum_i |\psi_i|^2 \right)^2}
\mbox{ . }
\end{equation}
%
%
This quantity reaches a non vanishing constant value in the large 
system size limit for localized states
whereas it behaves as $1/N_s$ for extended states, $N_s$ being the 
total number of sites.
For each value of $\sigma$, we have averaged over $10^2$ realizations 
of the disorder and computed the
average value of IPR over the whole energy spectrum. The results are 
plotted in Fig.~\ref{IPR} for several
values of the reduced flux, as a function of the disorder strength $\sigma$.
%
%
%
%

We first notice that, for $f=1/2$ the averaged IPR is indeed almost 
independent of $\sigma$ up to
$\sigma\simeq 1$, in agreement with the qualitative picture given 
before. As $\sigma$ is further increased,
the averaged IPR increases which is interpreted as a relative 
relocalization associated to an interband
mixing. For $f<1/2$, another energy scale $D(f)$ related to the mean 
band width in absence of
disorder emerges. If $\sigma<D(f)$, the localization length is finite 
for any $\sigma$ and decreases with
$\sigma$ as indicated, for instance at $f=0.495$, by the 
corresponding increase of the IPR for low $\sigma$. This
lasts until $D(f)\lesssim \sigma \lesssim 1$ where eigenstates no 
longer evolves with $\sigma$. The
intermediate regime in smoothly connected to the plateau observed for 
$f=1/2$. It
is, in some sense, what remains of Aharonov-Bohm cages as the field 
is no longer fixed to its special value
$f=1/2$. Finally, the strong disorder regime ($\sigma \gtrsim 1$), 
where the localization length becomes
comparable to the lattice spacing, is found not to depend sensitively 
of $f$ which sounds quite reasonable.

\begin{figure}[h]
\centerline{\epsfxsize=95mm
\epsfbox{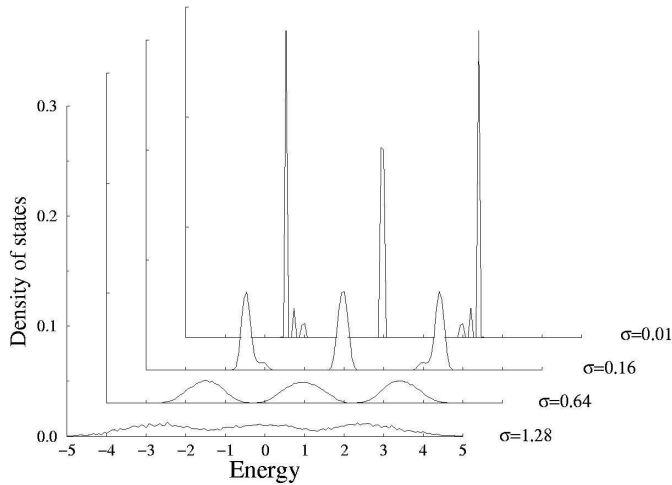}}
\caption{Variation of the averaged density of states at $f=1/2$ for 
various disorder strength
$\sigma$ ($N_s=605$ sites).}
\label{Histo}
\end{figure}

Similar conclusions can be drawn by looking at the density of states. 
Plots of this quantity at $f=1/2$ for
various values of $\sigma$ are shown in Fig.~\ref{Histo}. Note that 
the splitting of the three main peaks
for $\sigma=0.01$ is due to edge states. The peaks broaden linearly 
with $\sigma$
in the plateau region which clearly ends when they merge.

Finally, we would like to mention a related study concerning the 
influence of the disorder on
the transmission properties of quantum 
networks\cite{Vidal_Transmission}. For sufficiently
weak disorder, it has been shown that the periodicity with respect to 
the magnetic flux of the magnetoresistance
remains equals to $h/e$ in the ${\cal T}_{3}$ lattice whereas it is 
completely dominated by the weak localization
regime ($h/2e$) for more conventional structures, e.~g. the square 
lattice. This indicates that the Aharonov-Bohm
cages resists to a small amount of disorder. We emphasize that 
beautiful experiments by Naud {\it et al.}
have recently shown up this phenomenon in two-dimensional electron 
gas (GaAs/GaAlAs)\cite{Naud_T3}, measuring the
magnetoresistance of mesoscopic artificial structures with the ${\cal 
T}_{3}$ geometry.

%
%
%
%
%
%

\section{The edges states}
\label{Edges}

%
%
As already mentioned, interesting edge state properties appear at 
$f=1/2$. We could study these states
explicitely for the tight-binding hamiltonian $H_0$ but, for the sake 
of comparing with the experimental
situations, we shall consider a continuous model where the network is 
made up of perfectly
conducting one-dimensional wires. The Schr\"odinger equation for a 
free particle of mass $m$ moving along a
wire reads~:
%
%
\begin{equation}
-{\hbar^2 \over 2 m} {\mbox{d}^2 \psi\over\mbox{d} x^2}=E \psi
\mbox{ , }
\end{equation}
%
%
where $x$ denotes the coordinate along the wire. For simplicity, we 
assume that no magnetic field is
present but we will introduce it as soon as we will need it.
If the extremities of the wire corresponds to nodes $i$ and $j$, and 
if the origin is chosen such that
$x=0$ for node $i$ and  $x=a$ for node $j$, we may write the general 
solution along the wire $(i,j)$ as~:
%
%
\begin{equation}
\psi^{ij}(x)={1 \over \sin (ka) }
\left[ \psi_i \sin (k(x-a))+ \psi_j \sin (x) \right]
\mbox{ ; } E(k)={\hbar^2 k^2\over 2 m}
\mbox{ . }
\end{equation}
%
%
For clarity, we have omit the index $k$, and we have set 
$\psi_i=\psi^{ij}(x=0)$ and $\psi_j=\psi^{ij}(x=a)$.
The  amplitudes on the nodes are constrained by a linear boundary 
conditions at each node of the network. A
frequently used condition which is compatible with current 
conservation at the node $i$ is~:
%
%
\begin{equation}
\sum_{\langle i,j \rangle} \left.{\mbox{d} \psi^{ij}(x)\over\mbox{d} 
x}\right|_{x=0}=0
\mbox{ . }
\end{equation}
%
%
Here, the sum is taken over the $z_i$ wires emerging from the node 
$i$. If we also impose the continuity
condition of the wave function at each node ($\psi^{ij}(x=0)$ 
independent of $j$), we obtain the following
set of equations~:
%
%
\begin{equation}
\lambda z_i \psi_{i}=\sum_{\langle i,j \rangle} \psi_j
\label{lameq}
\mbox{ . }
\end{equation}
%
%
where we have introduced $\lambda=\cos k a$. So, a single value of 
$\lambda$ generates an infinite discrete
family of eigenstates with energy
$E(k_{n,\lambda})={\hbar^2 \over 2 m} \left( k_\lambda+{2\pi n\over 
a} \right)^2$ provided
$\lambda=\cos k_\lambda a$.
To describe each eigenstate only once, we may impose $k$ to lie in 
the interval $[-\pi/a, \pi/a]$. Note
that Eq.~(\ref{lameq}) cannot be directly interpreted as an 
eigenvalue equation for a tight-binding problem
since, in the general case,  $z_i$ varies from one site to another. 
For an infinite ${\cal T}_{3}$ lattice, or a finite
one with appropriate boundary conditions, the coordination number 
$z_i$ can only be equal to 3 or 6. It is
therefore possible to perform the same decimation of the three 
fold-coordinated sites as for the
tight-binding problem to obtain the same equation as 
Eq.~(\ref{spectre1d}) by setting
$\varepsilon^2=18\lambda^2$. This simple correspondence between both 
problems also holds in the presence of
a uniform magnetic field. In zero field, since
$\varepsilon$ runs from $-\sqrt{18}$ to $\sqrt{18}$, $\lambda$ runs 
from $-1$ to $+1$ and therefore
$k_\lambda$ from $[-\pi/a, \pi/a]$. So, there is no gap in the free 
electron spectrum on the network.
As soon as the magnetic field is present, $k_\lambda$ has to be chosen from
$[-\pi/a+\delta/a,-\delta/a] \cup [+\delta/a,\pi/a-\delta/a]$ where
$\cos(\delta a)=\varepsilon_{max.}(f) / \sqrt{18}$, 
$\varepsilon_{max.}(f)$ denoting the largest eigenvalue
of $H_0$ at reduced flux $f$. Thus, the free electron spectrum 
exhibits an infinite number of non
overlapping bands, for generic $f$. Gaps in the spectrum have two 
different origins.Most of them
corresponds to change the internal motion of the particle along a 
single link, which amounts to turn $n$
into $n+1$. Other gaps are induced by the magnetic field, and already 
appears as gaps in the set of allowed
$\lambda$ values ($\lambda$ spectrum). For the special value $f=1/2$, 
the cage effect is manifested by the
presence of a pure point-spectrum at energies
$E_n={\hbar^2 \over 2 m a^2} \left(\pm \pi/2 \pm \alpha +2\pi n 
\right)^2$, with
$\alpha=\arcsin (1/\sqrt{3})$. Each of these levels is highly 
degenerate, reflecting all the possible cage
states in the network.

\begin{figure}[h]
\centerline{\epsfxsize=50mm
\epsfbox{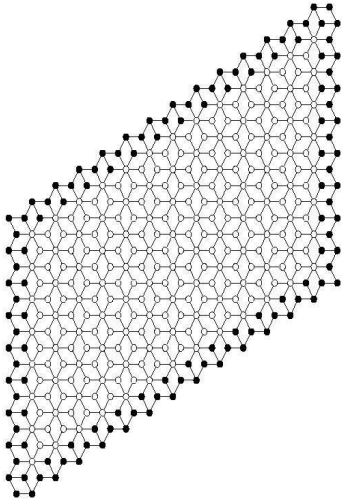}}
\caption{An example of edge states in a ${\cal T}_{3}$ wire network. 
We have put a black circle on each
node  for which the wave function has a non vanishing amplitude.}
\label{Edgesfig}
\end{figure}

\begin{figure}[h]
\centerline{\epsfxsize=100mm
\epsfbox{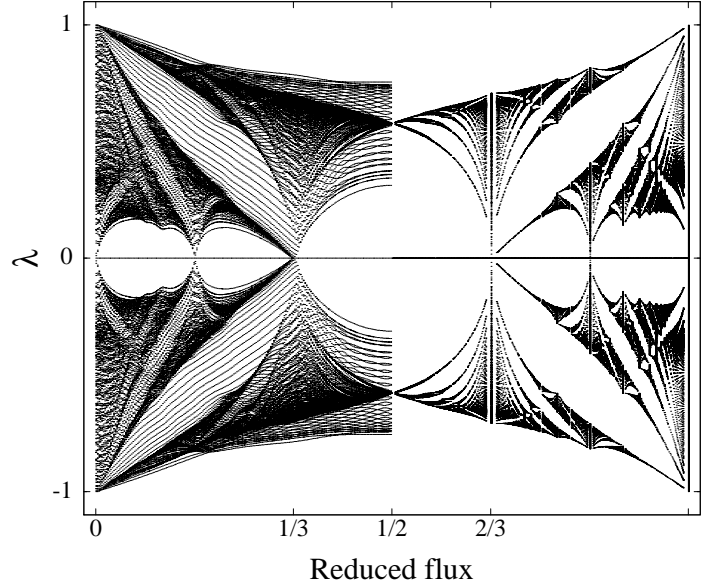}}
\caption{Comparison of the $\lambda$ spectrum for the infinite 
lattice (right) and the finite
size lattice displayed in Fig.~\ref{Edgesfig} (left).}
\label{pap2}
\end{figure}

Since some experiments have been performed on superconducting wire 
networks, it is worth recalling briefly
the connection between the $\lambda$ spectrum and the estimate of the 
superconducting critical temperature as
a function of the external flux $f$. At the transition, the order 
parameter builds up from the propagating
modes of the linear Ginzburg-Landau equation (written here in zero 
field for simplicity)
along each wire\cite{Alexander_fils_supra,DeGennes_fils_supra}~:
%
%
\begin{equation}
-{\mbox{d}^2 \psi\over\mbox{d} x^2}+{1 \over \xi^2} \psi=0
\mbox{ . }
\end{equation}
%
%
The critical temperature is obtained from the largest value of $\xi$ 
denoted by  $\xi^*$, for which
eigenvalues exist. $\xi^*$ is known as the Ginzburg-Landau 
macroscopic coherence length and we have
$T_c(f)-T_c(0) \propto 1/ (\xi^*)^2$. To determine $\xi^*$, we follow 
the same procedure as for the free
particle spectrum. We then have  $(k^*)^2 =(\xi^*)^2$ with $k^*\in 
[0,\pi/a]$ and $\cos
(k^*a)=\lambda_{max.}(f)$. For an infinite lattice, one has~:
$\lambda_{max.}(f)=\varepsilon_{max.}(f)/\sqrt{18}$ so that the 
$T_c(f)$ curve is directly related to the edge
of the tight-binding spectrum. This mapping has led to accurate 
comparisons between this simple model
and the experimental data\cite{Abilio_T3}. However, for free 
boundaries system, such a simple correspondence
no longer holds, since the coordination number can be different from 
3 and 6 on the edges. For these reasons,
our discussion of edge states will be given mostly in term of the 
linear problem (\ref{lameq}) rather than
in the tight-binding language. The main striking result is the 
appearance of very sharp edge states at
$f=1/2$. For those states, the amplitudes $\psi_i$'s are vanishing on 
most of the sites except on a
finite width strip concentrated near the boundaries (see in 
Fig.~\ref{Edgesfig}).
%
%
%
%
For any shape of the boundary, these edge states are always 
dispersive (along the edges). Note
that in the tight-binding version, non dispersive  edge states can 
also appear depending on the
boundary shape. Another qualitative difference is that, in the 
tight-binding problem,  the energy
of these states always appears inside the main gaps of the infinite 
lattice spectrum whereas in
the wire network, the $\lambda$ spectrum  exhibits edge states with 
energy outside the bulk
spectrum. We have displayed in Fig.~\ref{pap2} the spectrum of the 
finite lattice shown in Fig.~\ref{Edgesfig}
and the infinite lattice spectrum.
%
%
%
%
The dispersive edge states are clearly visible. These have important 
consequences for
superconducting networks since they offer the possibility to nucleate 
superconductivity
on the edges of the sample, at temperatures slightly above the bulk
$T_c(f)$\cite{DeGennes_supra,Bezryadin}.

\begin{figure}[h]
\centerline{\epsfxsize=60mm
\epsfbox{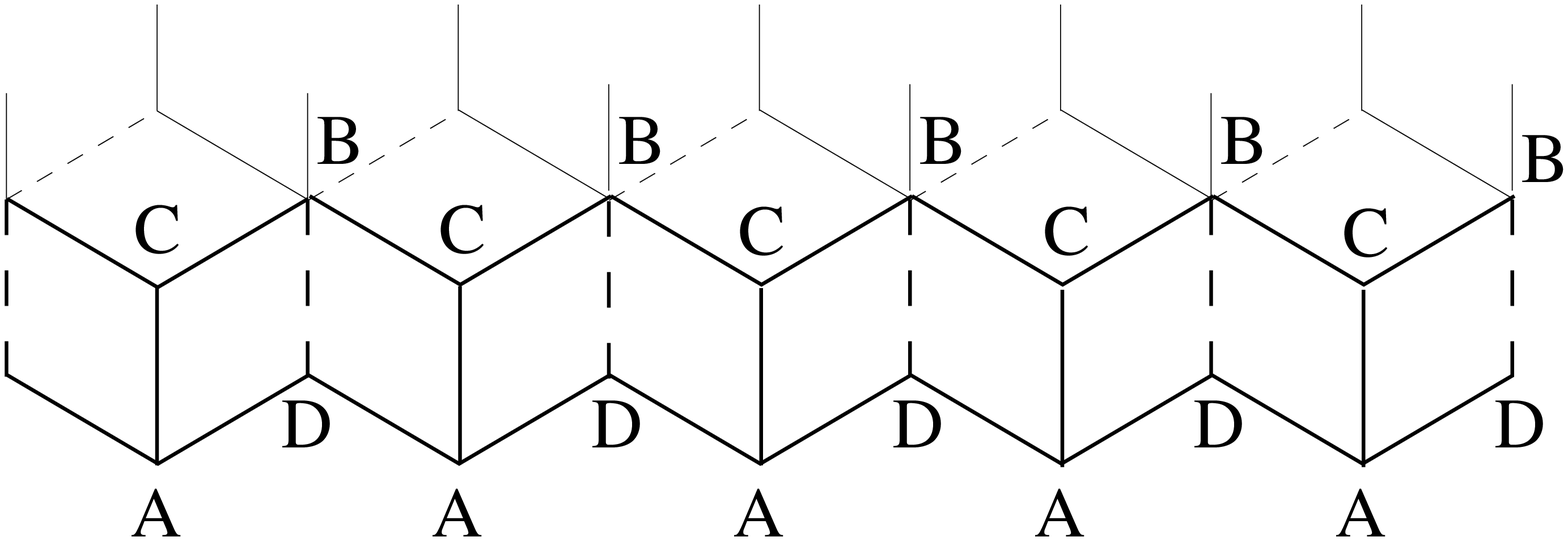}}
\vspace{5mm}
\caption{A possible edge choice for the semi-infinite ${\cal T}_{3}$ 
network (same convention as in
Fig.~\ref{Star4gauge}).}
\label{semiT3}
\end{figure}

Notice that for specific boundaries, the dispersive edge states can 
run all around the sample as illustrated in
Fig.~\ref{Edgesfig}, providing possibly an interesting example of a 
non chiral (quasi) one-dimensional conductor.
Some experiments on ballistic networks etched in a two-dimensional 
electron gas in GaAs
heterostructures have given some indications that edge states may 
have some influence in
transport measurement\cite{Naud2}. Indeed, the variation of the 
conductance through such a ${\cal T}_{3}$ network with
external field seems to depend on the geometry of the current 
pattern. On some samples, a
stronger dip in the conductance at $f=1/2$ has been observed when the 
current is injected
directly in the bulk, in comparison to the usual setup where the 
current is injected and
collected on the edges.  In this latter configuration, a non 
vanishing conductance
may be found for a perfect system thanks to these propagating edge 
states\cite{Vidal_Transmission}.
For illustration, we give an analytical expression for edge states 
($\lambda$ spectrum) coresponding to the
semi-infinite ${\cal T}_{3}$ network displayed in Fig.~\ref{semiT3}.
For these specific edges, the only non vanishing amplitudes are $A, 
C$ and $D$ ($B=0$).
Therefore, we may choose a convenient gauge in the edge region as 
displayed in  Fig.~\ref{semiT3}.
%
%
%
%
Assuming propagating waves in the edge direction with wave vector 
$k$, we must satisfy the
following set of equations~:
%
%
\begin{eqnarray}
3 \lambda A&=& C+ 2 D\cos (ka) \\
0&=& 2 C\cos (ka)-D \\
3 \lambda C&=& A \\
3 \lambda D&=& 2A \cos (ka)
\mbox{ . }
\end{eqnarray}
%
%
The corresponding $\lambda$ spectrum is given by $\lambda=\pm {1\over 
3} \sqrt{1+4\cos^2(ka)}$.
This example shows a dispersive edge spectrum $\lambda \in 
[1/3,\sqrt{5}/3]$ that lies below and
above $\lambda=1/\sqrt{3}$ which is the maximum value of $\lambda$ 
for the infinite system. As
previously discussed, these states are thus more favourable than the 
bulk states for the nucleation
of superconducting regions in a wire network upon cooling from high 
temperature.

\begin{figure}[h]
\centerline{\epsfxsize=100mm
\epsfbox{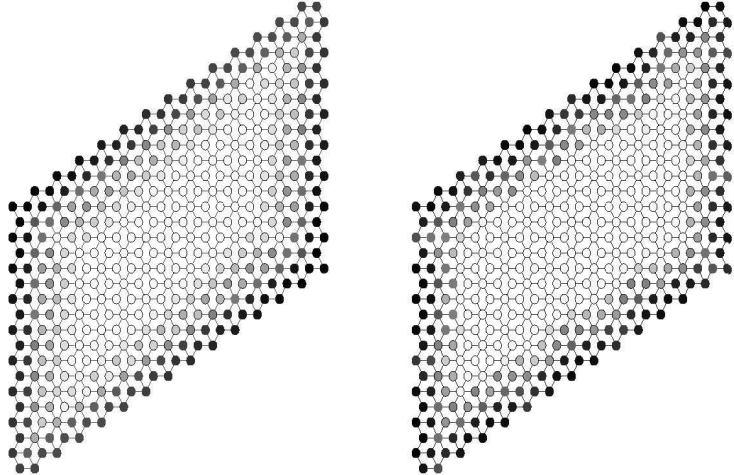}}
\caption{An example of edge states for $f=0.52, \sigma=0$ (left) and 
$f=1/2, \sigma=0.01$ (right). The grey
level on site $i$ is proportional to the $\log|\psi_i|$.}
\label{away}
\end{figure}

We have also investigated how these very narrow edge states evolve as 
some perturbations
away from the pure system at $f=1/2$  are gradually introduced. On 
Fig.~\ref{away}, some edge
states are represented for a pure system ($\sigma=0$) at $f=0.52$ and 
for a disordered system
($\sigma=0.01$) at $f=1/2$.  Both cases are very similar.  The main 
feature is the spatial broadening of the
wave function towards the center of the sample. However, we observe a 
rather smooth evolution of these states
as the strength of the perturbation is increased.
%
%
%
%

Up to now, we have always consider that the electrons were 
independent. In the last section, we shall try to analyze
the importance of the interactions between particles, in the ${\cal 
T}_{3}$ lattice.
Note that for simpler systems, the two interacting particle problem under magnetic field has already revealed some
unusual interesting features\cite{Barelli_TIP}.

%
%
%

\section{The interacting case}
\label{Interactions}

%
%
%

Of course, the many-body problem is certainly one of the most 
difficult to tackle. Here, our goal is to
study, in a simple approach, the effect of the electron-electron 
interactions on the
Aharonov-Bohm cages. We underline that this type of system provides a 
good starting point to
understand the competition between localization and interaction even 
if the nature of the confinement
is due to the magnetic field and not to disorder.  In our case, the 
main departure from a realistic
disordered model is indeed the preservation of the translation 
invariance in a localized system.
In the following, we first focus on the two electron problem that 
already displays very interesting
features, and we fix the reduced flux to its ``critical value" 
$f=1/2$. The system is described
by the following Hubbard hamiltonian~:
%
%
\begin{eqnarray}
H_{Hub.}&=&\sum_{\langle i,j \rangle, \sigma=\uparrow,\downarrow}  t_{ij} \:
c^\dagger_{i, \sigma}\,c_{j, \sigma}+U \sum_{i} n_{i, 
\uparrow}\,n_{i, \downarrow}
\label{ham.hubbard}\\
&=& H_0 + H_{int.}
\mbox{ , }
\end{eqnarray}
%
%
where $c^\dagger_{i,\sigma}$ and  $c_{i,\sigma}$ denotes the creation 
and annihilation operator of a fermion with spin
$\sigma$ respectively,
$n_{i,\sigma}=c^\dagger_{i,\sigma}\,c_{i,\sigma}$ the  density of 
spin $\sigma$ fermion on site $i$,
and $\langle \ldots \rangle$ stands for nearest neighbor pairs.
Note that, since the particles considered here are fermions, the 
interaction term $U$ is only efficient in
the singlet sector where the orbital part of the wave function is 
symmetric. For simplicity, we completely neglect
the coupling between the magnetic field and the electron spin.

Here, we shall mainly discuss the case of the ${\cal T}_{3}$ lattice 
but we must mention the work presented in
ref.\cite{Vidal_Chapelet} on the chain of loops displayed in 
Fig.~\ref{chap}.  For this system, we have
shown that a Hubbard-like interaction term leads to a delocalization 
which is directly related to the
emergence of dispersive bands in the two-particle spectrum. Of 
course, such a study is much more complex in
the ${\cal T}_{3}$ lattice, but, as we shall see, it is however 
possible to show that the same effect is
present in this two-dimensional structure.

\begin{figure}[h]
\centerline{\epsfxsize=65mm
\epsfbox{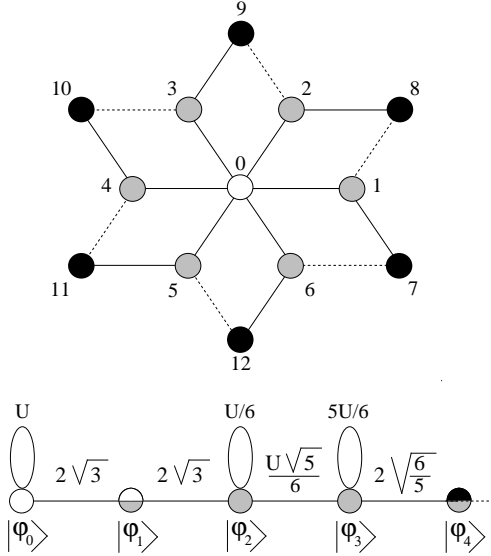}}
\caption{Aharanov-Bohm cage of a two-particle state initially 
localized on the central site, and its recursion chain
(same convention as in Fig.~\ref{Star4gauge}).
For $U\neq 0$, the black sites are reached.}
\label{Recursion2e}
\end{figure}

First, let us remark that if, at a given time, both electrons are far 
from each other, the local
particle density is non vanishing in a cage that is simply the
superposition of each individual cages. The interesting case thus 
arises when the electrons get close.
We consider, for instance, an initial state  where both electrons 
$\uparrow$ and $\downarrow$ are located
on the same six-fold coordinated site, and compute the coefficients 
of the recursion chain for this two-particle
wave function. With the site numbering given in 
Fig.~\ref{Recursion6}, we denote this initial state by
$|\varphi_0\rangle=|0,0\rangle$. More generally,  a ket $|i,j\rangle$ 
represents a state where the particle
with spin $\uparrow$ is located on site  $i$ and the particle with 
spin $\downarrow$ is located on
site $j$. Note that we must, in principle, work with the symmetrized 
ket but, as shown below, since our initial ket
$|0,0\rangle $ is already symmetric, it will automatically generate 
symmetrized kets.
Applying the recursion algorithm in the gauge displayed in 
Fig.~\ref{Recursion2e}, one obtains~:
%
%
\begin{eqnarray}
H_{Hub.}|\varphi_0\rangle&=& U |0,0\rangle+\sum_{i=1}^6 |i,0\rangle+|0,i\rangle
\\
\Rightarrow \:\:\:  a_0&=&U \, , \: b_1=2\sqrt{3} \, , \:
|\varphi_1\rangle= {1 \over 2\sqrt{3}}\sum_{i=1}^6 |i,0\rangle+|0,i\rangle
\mbox{,}\\
H_{Hub.}|\varphi_1\rangle&=&{1 \over \sqrt{3}} \sum_{i,j=1}^6 |i,j\rangle
+ b_1 |\varphi_0\rangle
\\
\Rightarrow \:\:\:  a_1&=&0  \, , \: b_2=2 \sqrt{3} \, , \:
|\varphi_2\rangle= {1 \over 6}\sum_{i,j=1}^6 |i,j\rangle
\mbox{,}\\
H_{Hub.}|\varphi_2\rangle&=& {U \over 6} \sum_{i=1}^6 |i,i\rangle
+ b_2 |\varphi_1\rangle
\\
\Rightarrow \:\:\:  a_2&=&{U \over 6} \, , \: b_3={U \sqrt{5} \over 6} \, , \:
|\varphi_3\rangle= {1 \over \sqrt{5}}\left({5 \over 6}\sum_{i=1}^6 |i,i\rangle
-{1\over 6} \sum_{i\neq j=1}^6 |i,j\rangle \right)
\mbox{,}\\
H_{Hub.}|\varphi_3\rangle&=& {1 \over \sqrt{5}}
\left( {5 \, U \over 6} \sum_{i=1}^6 |i,i\rangle +
  \sum_{i=1}^6 |i,i+6 \rangle - |i,i+7 \rangle +|i+6,i \rangle
-|i+7,i \rangle \right)
\\
\Rightarrow \:\:\:  a_3&=&{5 \,U \over 6}  \, , \:  b_4=2\, \sqrt{6 
\over 5} \, , \:
|\varphi_4\rangle={1 \over 2 \sqrt{6}}  \sum_{i=1}^6 |i,i+6 \rangle - 
|i,i+7 \rangle +|i+6,i \rangle -|i+7,i \rangle
\mbox{.}
\end{eqnarray}
%
%
Since $|\varphi_4\rangle$ has a non vanishing amplitude on the second 
shell sites ($i\geq 7$) the two-particle wave
packet can spread, so that the Aharonov-Bohm cages associated to 
$|0,0\rangle$ seems to be unbound or, at least, larger
than the one obtained for $|0\rangle$ at $U=0$. Actually, we have 
checked numerically that the $b_{i\geq 5}$'s are non
vanishing so that the cage is really unbound.
%
%
%
%

As for the chain of loops, the propagation is made possible by the 
existence of two-body bound states in the
spectrum. The same analysis as in ref.\cite{Vidal_Chapelet} could be 
done in the ${\cal T}_{3}$ lattice but the
two-dimensional character of this later structure makes it more 
difficult. Nevertheless, if one focuses on
the small interaction case ($U \ll t$) it is possible to show the 
emergence of dispersive bands near the
ground state energy. Indeed, one  can, in this case, treat $U$ as a 
perturbation on the
infinitely degenerate level $2 \varepsilon_-$ without considering the 
higher energy levels.

\begin{figure}[h]
\centerline{\epsfxsize=100mm
\epsfbox{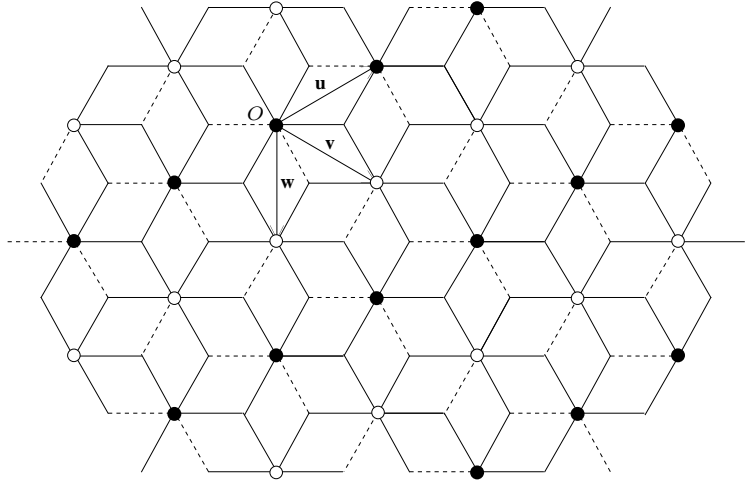}}
\caption{
A possible gauge choice for the ${\cal T}_{3}$ lattice at  $f=1/2$ 
(same convention as in Fig.~\ref{Star4gauge}).}
\label{Star3gauge}
\end{figure}

\begin{figure}[h]
\centerline{\epsfxsize=120mm
\epsfbox{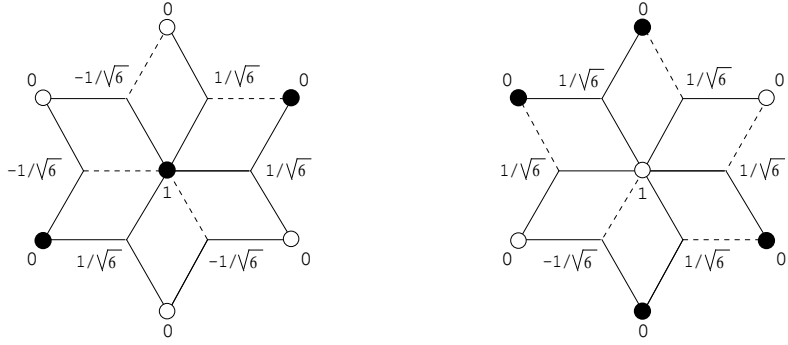}}
\caption{Non-normalized one-particle cage eigenstates associated to 
the two types of 6-fold
coordinated sites (see Fig.~\ref{Star3gauge}). The amplitude on each 
site is given.}
\label{cagebasis}
\end{figure}

To solve this problem, we shall use the periodic gauge displayed in 
Fig.~\ref{Star3gauge} that
provides real hopping terms.
%
%
%
%
Remark that the periodicity of the structure has doubled and we must now
distinguish two different types of 6-fold coordinated sites denoted 
by $(\bullet)$ and $(\circ)$.
If one restricts the two-body problem analysis to the subspace 
corresponding to $2\varepsilon_-$
(for
$U=0$), it is convenient to build the two-particle state basis as the 
tensor product of the one-particle
orthogonal ``cage" basis displayed in Fig.~\ref{cagebasis}.
%
%
%
%
Since the problem is invariant under a translation of the center of 
mass of the two particles, the
basis of the singlet states sensitive to $U$ can be expressed in 
terms of the following Bloch waves~:
%
%
\begin{eqnarray}
|\varphi_0({\bf K}) \rangle&=& {1 \over \sqrt{N_s}} \sum_{{\bf r} \in 
\Lambda} e^{i{\bf K.r}}
|{\bf r}, {\bf r}\rangle\\
|\varphi_1({\bf K}) \rangle&=& {1 \over \sqrt{N_s}} \sum_{{\bf r} \in 
\Lambda} e^{i{\bf K.r}}
|{\bf r+ v}, {\bf r+ v}\rangle\\
|\varphi_{\bf u_0}({\bf K}) \rangle&=& {1 \over \sqrt{N_s}} 
\sum_{{\bf r} \in \Lambda} e^{i{\bf K.r}}
|{\bf r}, {\bf r + u}\rangle_S\\
|\varphi_{\bf u_1}({\bf K}) \rangle&=& {1 \over \sqrt{N_s}} 
\sum_{{\bf r} \in \Lambda} e^{i{\bf K.r}}
|{\bf r + v }, {\bf r + v + u}\rangle_S\\
|\varphi_{\bf \pm v}({\bf K}) \rangle&=& {1 \over \sqrt{N_s}} 
\sum_{{\bf r} \in \Lambda} e^{i{\bf K.r}}
|{\bf r}, {\bf r \pm v}\rangle_S\\
|\varphi_{\bf \pm w}({\bf K}) \rangle&=& {1 \over \sqrt{N_s}} 
\sum_{{\bf r} \in \Lambda} e^{i{\bf K.r}}
|{\bf r}, {\bf r \pm w}\rangle_S
\mbox{,}
\end{eqnarray}
%
%
where the origin $O =(0,0)$, ${\bf u}$, ${\bf v}$ and ${\bf w}$ are 
represented in Fig.~\ref{Star3gauge}~;
$\Lambda$ is the lattice generated by ${\bf u}$ and ${\bf v+w}$ with 
periodic boundary
conditions and containing $N_s$ sites~; ${\bf K}$ is a wave vector 
lying in the first Brillouin zone associated
to $\Lambda$. Note that the orthogonal basis used here to define the 
different vectors is different
than the basis used in section~\ref{Cages}.
 From now on, the ket $|{\bf r}, {\bf r'}\rangle_S$ represents the 
normalized space symmetric state in which
one electron is in a cage state (eigenstate of the one-particle non 
interacting problem) localized around
a six-fold coordinated site located in ${\bf r}$ and the other 
electron is in a cage state localized around
a six-fold coordinated site located in ${\bf r}'$. Because of the 
Pauli principle, these kets corresponds to
antisymmetric spin wave functions and therefore to singlet states.
Note that $|{\bf r}; {\bf r}\rangle$ is automatically symmetric so 
that we have omit the index $S$ for this
state.  Naturally, if $|{\bf r}-{\bf r}'|>\sqrt{3}$, the electrons 
do not interact so that one may only to
consider the 8 two-particle states defined above. Thus, in each 
irreducible representation indexed by ${\bf
K}$, one just has to diagonalize $H_{Hub.}$  in the subspace 
generated by these vectors. Furthermore, as shown in the
Appendix~\ref{Diago}, $H_{Hub.}$ can be block diagonalized into two 
isospectral matrices.
After shifting the energies so that for $U=0$ the eigenenergies are 
equal to zero (and not
to $2\varepsilon_-$), the eigenvalues (in $U$ units) are given by the 
roots of the following characteristic polynomial~:
%
%
\begin{equation}
P(\varepsilon,{\bf K})=\varepsilon
[10368\varepsilon^3-\varepsilon^2 (3888+288 \mu) +\varepsilon(258+ 20 
\mu + 2\nu)+ \mu -3]
\mbox{.}
\label{polynom}
\end{equation}
%
%
%
%
\begin{eqnarray}
\mu&=&\cos({\bf K}.{\bf u}) + \cos({\bf K}.{\bf v}) + \cos({\bf K}.{\bf w})\\
\nu&=&\cos[{\bf K}.{\bf (u+v)}] + \cos[{\bf K}.{\bf (2u-v})] 
+\cos[{\bf K}.{\bf (2v-u})]
\mbox{.}
\end{eqnarray}
%
%
Reminding that ${\bf w}={\bf v}-{\bf u}$, one can easily check that 
$P$ is unchanged by the tranformation
${\bf u}\leftrightarrow {\bf v}$, as it obviously should.

It is readily seen in Eq.(\ref{polynom}) that $\varepsilon=0$ is 
always an eigenvalue for all ${\bf K}$. The
associated eigenvectors have a vanishing amplitude on the kets 
$|\varphi_0({\bf K})
\rangle$ and $|\varphi_1({\bf K}) \rangle$ that corresponds to states 
where both electrons are in the same cage.
The absence of dispersion for this family of states suggests that 
they may be generated by a set localized
two particle singlets. As in the one particle problem (see Section 
\ref{Cages}), it is even possible to exhibit such
states where the center of mass is confined in a finite area. A 
possible choice is~:
%
%
\begin{equation}
|\psi_{Bound}\rangle={1\over \sqrt{6}}
\left(
|{\bf 0}, {\bf u} \rangle_S - |{\bf 0}, -{\bf u} \rangle_S -
|{\bf 0}, {\bf v} \rangle_S + |{\bf 0}, -{\bf v} \rangle_S +
|{\bf 0}, {\bf w} \rangle_S + |{\bf 0}, -{\bf w} \rangle_S
\right)
\end{equation}
%
%
The presence of these localized singlet bound states at zero energy 
seems to be a particular feature of this lattice.

\begin{figure}[h]
\centerline{\epsfxsize=90mm
\epsffile{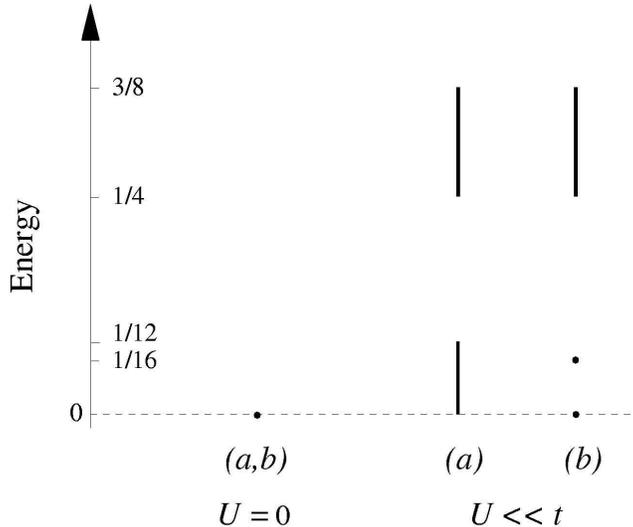}}
\caption{Perturbative low-energy two-particle spectrum for $(a)$ the 
${\cal T}_{3}$ lattice, $(b)$ the chain of loops.}
\label{spectrum2e}
\end{figure}

Apart from this non dispersive band, a close inspection of the 
polynomial $P$ allows one to show that when ${\bf
K}$ runs over an elementary cell of the reciprocal lattice, the roots 
of $P$ describe two separate dispersive
bands spreading from 0 to 1/12 for the lower one and from 1/4 to 3/8. 
The corresponding
eigenvectors are linear combinations of the 8 vectors introduced 
above and are actually extended Bloch
waves. These two-body bound states are delocalized and thus allow for 
a propagation of
two-particle wave functions as we have already noticed using the 
recursion analysis (see in
Fig.~\ref{Recursion2e}).  The same perturbative approach in the 
degenerate ground state
can be achieved for the chain of loops\cite{Vidal_Chapelet}.
In this latter case, the spectrum, for $U\ll t$ consists in two non 
dispersive bands with eigenenergies 0
and 1/16, and one dispersive band spreading also from $1/4$ to
$3/8$.
Here again these energies are measured in $U$ units and shifted such 
that, for $U=0$, the two-body ground
state corresponds to a zero energy. We have displayed in 
Fig.~\ref{spectrum2e} these two spectra.

Two important differences arise between these two structures. First, 
there is no non dispersive singlet bound
state at zero energy for the chain of loops, except the trivial ones 
involving two remote localized electron states.
Non dispersive singlet bound states do exist but their energy is 
strictly positive and depends on $U$.
Second, there is a gap between the $U=0$ ground state and the first 
dispersive band (bound state), by contrast to the
${\cal T}_{3}$ lattice.
%
%
%
%
The fact that all eigenenergies of the projected hamiltonian are 
positive is a simple consequence of the same
property for the interacting part $H_{int.}$. Denoting by $Q$ the 
projector on the lowest energy level of
$H_0$, the lowest order degenerate perturbation theory proposed above 
amounts to diagonalize $Q H_{int.} Q$.
For any two-particle eigenstate $|\psi\rangle$, we have $\langle \psi 
| H_{int.} | \psi \rangle\geq 0$ if
$U\geq 0$. In particular, this imposes  all eigenvalues of $Q 
H_{int.} Q$ to be positive (or vanishing).
 From this consideration, the appearance of non trivial (degenerate) 
zero energy bound states in the singlet
sector is quite remarkable.

Of course, it would be very interesting to characterize the 
properties of the system in the presence of a finite
density of particles. In the following discussion, we shall always 
assume $U\ll t$, so that we will only pay
attention to the projected hamiltonian $Q H_{int.} Q$ which is, as 
discussed above, a positive operator.
Denoting by $N_e$ the total number of electrons and by $N_s$ the 
total number of sites, we introduce the density
$\rho=N_e/N_s$ that ranges from 0 to 2 because of the spin 
degeneracy.  In the non interacting case, for $0\leq \rho
\leq 2/3$, it is possible to put all the electrons, in the ground 
state of $H_0$. Thus, we will restrict our analysis
to this interval.

For $\rho \leq 1/3$, a possible choice for the ground state is to put 
each electron in a localized cage state
(around 6-fold coordinated sites). In this case, the energy per 
particle is simply $\varepsilon_-$.
At low density, if all occupied cages are further isolated from each 
other (no overlap), the interaction term
$H_{int.}$ does not affect these configurations which are therefore 
ground states of $Q H_{Hub.} Q$. These
configurations have a huge spin degeneracy ($2^{N_e}$). One can also
build ground state configurations containing some connected clusters 
of singly occupied cage states, provided
these clusters are fully spin polarized. Obviously, the total spins 
of these clusters may be chosen independently
from each other without any energy cost.
Note that additional configurations with the same energy may also be 
obtained using the zero energy two
particle localized singlet bound states discussed above.
An interesting issue would be to determine whether this class of ground
states exhausts all the possible ones, for $\rho \leq 1/3$.
Note that if the density $\rho$ lies beyond $\nu_T/3$ where $\nu_T$ is the
site percolation density of the triangular lattice which is formed by 
the 6-fold coordinated sites, we have an
infinite percolating cluster which leads to an infinite magnetization.

As $\rho$ approaches $1/3$, the ground state thus becomes more and 
more polarized. If $N_s$ denotes the total
number of sites, the problem of finding the ground state of the 
projected hamiltonian becomes more interesting
if $1/3 < \rho \leq 2/3 $. A simple class of eigenstates of $Q 
H_{Hub.} Q$ is obtained if the
magnetization is maximal. It corresponds to a total spin 
$S_{max.}=N_s/3 - N_e/2$. Indeed, because of the spin
rotation invariance, such states can be built from states with 
$N_\uparrow=N_s/3$ and
$N_\downarrow=N_e/2 - N_s/3$ after performing a global spin rotation. 
Here, $N_\uparrow$ and $N_\downarrow$
denotes the total number of electrons with up and down spins 
respectively. The condition $N_\uparrow=N_s/3$
means that all the single particle ground states of $H_0$ are fully 
occupied by $\uparrow$ spin electrons. For these
states, the $\downarrow$ spin electrons behave as completely 
localized free fermions with an individual excitation
energy equal to $U$. To show this, we denote by $\varphi_{\bf 
r}(i)=\langle i | {\bf r}\rangle$ the amplitude at site
$i$ of the state $| {\bf r}\rangle$ which is the single particle cage 
eigenstates of $H_0$ localized around the
six-fold coordinated site located at site ${\bf r}$, and we write~:
%
%
\begin{equation}
Q H_{int.} Q=\sum_{{\bf r}_1,{\bf r}_2,{\bf r}_3,{\bf r}_4}
u({\bf r}_3,{\bf r}_4|{\bf r}_1,{\bf r}_2) \:\:
c^\dagger_{{\bf r}_3, \uparrow} c^\dagger_{{\bf r}_4, \downarrow}
c_{{\bf r}_2, \downarrow} c_{{\bf r}_1, \uparrow}
\mbox{,}
\label{Hubbardbis}
\end{equation}
%
%
with~:
%
%
\begin{equation}
u({\bf r}_3,{\bf r}_4|{\bf r}_1,{\bf r}_2)=U \sum_j \varphi^*_{{\bf 
r}_3}(j) \varphi^*_{{\bf r}_4}(j)
\varphi_{{\bf r}_2}(j) \varphi_{{\bf r}_1}(j)
\mbox{.}
\end{equation}
%
%
In expression (\ref{Hubbardbis}), the fermions operator 
$c^\dagger_{{\bf r}, \sigma}$ and $c_{{\bf r}, \sigma}$
create and destroy an electron with spin $\sigma$ in a	state $| {\bf 
r}\rangle$.
On a state where all the single particle ground states of $H_0$ are 
fully occupied by up spins electrons,
$c^\dagger_{{\bf r}_3, \uparrow} c_{{\bf r}_1, \uparrow}$ acts as 
$\delta_{{\bf r}_3,{\bf r}_1}$ times the identity
operator. It is then easy to check that~:
$\sum_{{\bf r}} u({\bf r}_,{\bf r}_4|{\bf r},{\bf r}_2)= U 
\delta_{{\bf r}_4,{\bf r}_2}$. As a consequence, for any
state $| \psi \rangle$ such that $N_\uparrow=N_s/3$, we have
$Q H_{int.} Q | \psi \rangle=U N_\downarrow | \psi \rangle$. At this 
stage, it not clear whether the ground state of the
projected hamiltonian has the maximal value $S_{max.}$ of the total 
spin for $1/3 < \rho \leq 2/3 $. Nevertheless,
since this is true for $\rho=1/3$ according to the previous 
paragraph, this generalization is plausible. The main
question to be adressed is whether a $\downarrow$ spin may form a 
bound state with one or several ``magnons". By magnon,
we refer here to particle-hole like excitations which destroy one 
$\uparrow$ spin electron and create one $\downarrow$
spin electron. We leave this interesting but more complex problem for 
future investigations.

To conclude this section, we may say that in spite of the presence of 
very low lying dispersive
bound states for two particles, it is not so easy to turn the finite 
density system into a good
conductor. However, we emphasize that some qualitative changes may 
arise for finite values of
$U$, since a sizeable virtual occupancy of excited levels of $H_0$ 
becomes then possible.

\section{Conclusion and perspectives}

In this paper, we have studied three types of perturbations which 
affect the physical properties of
a two-dimensional lattice embedded in a magnetic field that presents 
Aharonov-Bohm cages. Two of them, namely a weak
Anderson-like disorder and the finite size effects do not drastically 
modify the main features of the ideal model. In
particular, if the disorder is not too important, the single particle 
energy eigenstates remain strongly localized for
half-integer fluxes per elementary tile, and this independently of 
the disorder strength. This lasts until the disorder
does not introduce a strong mixing between distinct degenerate levels 
of the unperturbed tight-binding hamiltonian.
By contrast, for generic values of the magnetic field, the amount of 
disorder-induced localization
is very sensitive to the disorder strength.

A more delicate situation is obtained in the presence of 
electron-electron interactions for the many-body system. We
have shown how they partially destroy the single particle 
localization by the generation of extended two
particle eigenstates. However, it is not clear at the present stage 
whether this mechanism is sufficient to induce a
metallic behavior for finite electron densities and half a flux 
quantum per tile. This important question clearly
deserves further investigations for the Hubbard model (studied here) 
and also  for more realistic model of continuous
narrow wires (few conducting channels), in connection to the 
experiments on two-dimensional electron gas. We expect to
find partially spin polarized ground states in a finite interval of 
electron density, and a fully polarized state when
the number of electron per site equals $1/3$. For slightly larger 
densities, an important question is to understand if
the system create some spin textures resembling to skyrmions in 
quantum Hall effect for filling fractions close to odd
integers.

In this context, it would be very interesting to study interacting 
hard core bosons on this lattice,
which would require numerical diagonalization on finite size 
clusters. Indeed, this model is closely related to the
physics of Josephson junction arrays. In the limit of small 
capacitance islands quantum fluctuations of the order
parameter phase are strong and the corresponding trend to localize 
Cooper pairs could be enhanced by the cage effect.
In the other limit where the Josephson coupling dominates, it is not 
clear that the system at
half-integer fluxes will develop a finite stiffness for spatial 
gradients of the order parameter phase. In this
semi-classical regime, we expect a rather large ground state 
degeneracy as suggested by vortex decoration experiments
on superconducting networks.

\acknowledgments

We would like to thank C.~C. Abilio, G. Faini, M.~V. Feigel'man, D. 
Mailly, P.  Martinoli,
G.  Montambaux, \mbox{C. Naud}, B. Pannetier, and E. Serret for 
fruitful and stimulating discussions.

\appendix
%
%
%

\section{Diagonalization of the two-electron problem}
\label{Diago}

%
%

As explained in section~\ref{Interactions}, the non trivial part of 
the hamiltonian is given, in each
irreducible representation indexed by ${\bf K}$, by the subspace 
generated by the eight vectors~:
$$\left( |\varphi_0({\bf K})\rangle,|\varphi_{\bf + v}({\bf 
K})\rangle, |\varphi_{\bf - v}({\bf K})\rangle
|\varphi_{\bf u_0}({\bf K})\rangle, |\varphi_{\bf u_1}({\bf K})\rangle,|
\varphi_{\bf + w}({\bf K})\rangle,|\varphi_{\bf - w}({\bf K})\rangle, 
\varphi_1({\bf K})\rangle \right).$$
In this basis,  the shifted hamiltonian $H'({\bf K})=(H_{Hub.} ({\bf 
K})-2\varepsilon_-I_8)/U$
($I_8$ denotes the $8 \times 8$ identity matrix) writes~:
%
%
\begin{equation}
H'({\bf K})=\left(
\begin{array}{cccccccc}
{7\over 24}+{\cos({\bf K}.{\bf u})\over 36} & {e^{-i{\bf K}.{\bf u}} 
\over 72 \sqrt{2}} &
-{e^{i{\bf K}.{\bf u}}\over 72 \sqrt{2}} & 0 &
{e^{-i{\bf K}.{\bf u}}- e^{-2i{\bf K}.{\bf v}}\over 72 \sqrt{2}} & 
{e^{i{\bf K}.{\bf u}}\over 72 \sqrt{2}}&
{e^{-i{\bf K}.{\bf u}} \over 72 \sqrt{2}} &
{e^{-i{\bf K}.{\bf v}}(\cos({\bf K}.{\bf v})+\cos({\bf K}.{\bf 
w}))\over 36}\\\\
&{1\over 36} & 0 & {1\over 72} &-{e^{-i{\bf K}.{\bf u}} \over 72} &
{1+e^{i{\bf K}.{\bf u}} \over 72} & 0 & -{e^{-i{\bf K}.{\bf u}} \over 
72 \sqrt{2}} \\\\
&&{1\over 36} & {e^{-i{\bf K}.{\bf u}} \over 72} & -{e^{-2i{\bf 
K}.{\bf v}} \over 72} & 0 &
{-1-e^{-i{\bf K}.{\bf u}} \over 72} & {e^{-i{\bf K}.{\bf (v+w)}} 
\over 72 \sqrt{2}} \\\\
&&&{1\over 36} & 0 & {e^{i{\bf K}.{\bf u}} \over 72} & -{1 \over 72} &
{1-e^{-i{\bf K}.{\bf (v+w)}} \over 72 \sqrt{2}} \\\\
&&&&{1\over 36} &-{e^{i{\bf K}.{\bf u}} \over 72} &{e^{2i{\bf K}.{\bf 
v}} \over 72} &0\\\\
&&&&&{1\over 36} &0 &{-1\over 72 \sqrt{2}} \\\\
&&&&&&{1\over 36} &{-e^{-2i{\bf K}.{\bf v}} \over 72 \sqrt{2}} \\\\
&&&&&&&{7\over 24}+{\cos({\bf K}.{\bf u})\over 36}
\end{array}
\right)
\mbox{ . }
\end{equation}
%
%
Note that since $H'({\bf K})$ is hermitian, we have only given its 
upper part. Now, let us introduce the following
vectors~:
%
%
\begin{eqnarray}
|\varphi({\bf K})\rangle_\pm&=& {1 \over \sqrt{2}}
\left(|\varphi_0({\bf K})\rangle \pm e^{-i{\bf K.v}} |\varphi_1({\bf 
K})\rangle \right)\\
|\varphi_{\bf u}({\bf K}) \rangle_\pm&=& {1 \over \sqrt{2}}
\left(|\varphi_{\bf u_0}({\bf K})\rangle \pm e^{-i{\bf K.v}} 
|\varphi_{\bf u_1}({\bf K})\rangle \right)\\
|\varphi_{\bf v}({\bf K}) \rangle_\pm&=& {1 \over \sqrt{2}}
\left(|\varphi_{\bf +v}({\bf K})\rangle \pm e^{i{\bf K.v}} 
|\varphi_{\bf -v}({\bf K})\rangle \right)\\
|\varphi_{\bf w}({\bf K}) \rangle_\pm&=& {1 \over \sqrt{2}}
\left(|\varphi_{\bf +w}({\bf K})\rangle \pm e^{i{\bf K.w}} 
|\varphi_{\bf -w}({\bf K})\rangle \right)
\mbox{,}
\end{eqnarray}
%
%
that corresponds (up to a phase factor), to symmetric and 
antisymmetric combinations of the initials vectors.
The hamiltonian does not connect the subspace generated by
${\cal B}_1=(|\varphi({\bf K})\rangle_-,|\varphi_{\bf v}({\bf K}) 
\rangle_- |\varphi_{\bf u}({\bf K})\rangle_+,|
\varphi_{\bf w}({\bf K}) \rangle_+)$ and the subspace generated by
${\cal B}_2=(|\varphi({\bf K})\rangle_+,|\varphi_{\bf v}({\bf K}) \rangle_+,
|\varphi_{\bf u}({\bf K})\rangle_-,|\varphi_{\bf w}({\bf K}) \rangle_-)$.
In each subpace, $H'({\bf K})$ writes~:
%
%
\begin{equation}
H'_{{\cal B}_1}({\bf K})=\left(
\begin{array}{cccc}
{7 \over 24} +{\cos({\bf K}.{\bf u}) + \cos({\bf K}.{\bf v}) + 
\cos({\bf K}.{\bf w}) \over 36}&
{e^{-i{\bf K.u}} - e^{-i{\bf K.w}} \over 72 \sqrt{2}} &
{e^{-i{\bf K.u}} - e^{-2i{\bf K.v}} \over 72 \sqrt{2}} &
{e^{-i{\bf K.u}} - e^{i{\bf K.v}} \over 72 \sqrt{2}}  \\\\
&{1\over 36}& {-e^{-i{\bf K.u}} - e^{-i{\bf K.v}} \over 72} &
{-e^{i{\bf K.v}} - e^{i{\bf K.w}} \over 72}\\\\
&&{1\over 36}& {e^{i{\bf K.v}}+e^{2 i{\bf K.v}} \over 72}\\\\
&&&{1\over 36}
\end{array}
\right)
\mbox{ , }
\end{equation}
%
%
%
%
%
%
\begin{equation}
H'_{{\cal B}_2}({\bf K})=\left(
\begin{array}{cccc}
{7 \over 24} +{\cos({\bf K}.{\bf u}) - \cos({\bf K}.{\bf v}) - 
\cos({\bf K}.{\bf w}) \over 36}&
{e^{i{\bf K.(u+v)}} + e^{-i{\bf K.(v+w)}} \over 72 \sqrt{2}} &
{1 - e^{i{\bf K.(v+w)}} \over 72 \sqrt{2}} &
{-1 - e^{i{\bf K.(u+v)}} \over 72 \sqrt{2}} \\\\
&{1\over 36}& {e^{-i{\bf K.u}} - e^{-i{\bf K.v}} \over 72} &
{-e^{-i{\bf K.v}} - e^{-i{\bf K.w}} \over 72} \\\\
&&{1\over 36}& {e^{-i{\bf K.u}} - e^{-i{\bf K.w}} \over 72} \\\\
&&&{1\over 36}
\end{array}
\right)
\mbox{ . }
\end{equation}
%
%
%
The characteristic polynomia $P_{{\cal B}_1} (\varepsilon,{\bf K})$ 
and $P_{{\cal B}_2} (\varepsilon,{\bf K})$
associated to $H'_{{\cal B}_1}({\bf K})$ and $H'_{{\cal B}_2}({\bf K})$ respectively reads~:
%
%
\begin{eqnarray}
P_{{\cal B}_1}(\varepsilon,{\bf K})&=&{\varepsilon \over 10368}
[10368\varepsilon^3-\varepsilon^2 (3888+288 \mu_1) +\varepsilon(258+ 
20 \mu_1 + 2\nu_1)+ \mu_1 -3]\\
P_{{\cal B}_2}(\varepsilon,{\bf K})&=&{\varepsilon \over 10368}
[10368\varepsilon^3-\varepsilon^2 (3888+288 \mu_2) +\varepsilon(258+ 
20 \mu_2 + 2\nu_2)+ \mu_2 -3]
\mbox{,}
\end{eqnarray}
%
%
with~:
%
%
\begin{eqnarray}
\mu_1&=&\cos({\bf K}.{\bf u}) + \cos({\bf K}.{\bf v}) + \cos({\bf 
K}.{\bf w})\\
\mu_2&=&\cos({\bf K}.{\bf u})-\cos({\bf K}.{\bf v})-\cos({\bf K}.{\bf w})\\
\nu_1&=&\cos[{\bf K}.{\bf (u+v)}] + \cos[{\bf K}.{\bf (2u-v})] 
+\cos[{\bf K}.{\bf (2v-u})]\\
\nu_2&=&-\cos[{\bf K}.{\bf (u+v)}] - \cos[{\bf K}.{\bf (2u-v})] 
+\cos[{\bf K}.{\bf (2v-u})]
\mbox{ . }
\end{eqnarray}
%
%
In the orthogonal basis chosen in section \ref{Model} where ${\bf 
u}=a (3/2,\sqrt{3}/2)$, ${\bf v}=a (3/2,-\sqrt{3}/2)$
and ${\bf w}={\bf v}-{\bf u}$, it is straightforward to show that 
$P_{{\cal B}_1}(\varepsilon,{\bf K})= P_{{\cal
B}_2}(\varepsilon,{\bf K'})$ if ${\bf K}=(k_1,k_2)$ and ${\bf 
K'}=(k_1+\pi/3,k_2-\pi/ \sqrt{3})$.
Thus, it is  sufficient to analyze the roots of the polynomial
$P(\varepsilon,{\bf K})=P_{{\cal B}_1}(\varepsilon,{\bf K})$ for all 
${\bf K}$ belonging to an elementary cell of the
reciprocal lattice.


\end{document}